\documentclass[article]{JHEP3}

\usepackage{amsmath,epsf}
\usepackage{graphicx}
\usepackage{amsmath,epsfig}
\usepackage{amssymb,amsfonts}
\usepackage{latexsym}

\relax
\renewcommand{\theequation}{\arabic{section}.\arabic{equation}}

\def\be{\begin{equation}}
\def\ee{\end{equation}}

\newcommand{\bear}{\begin{eqnarray}}
\newcommand{\bea}{\begin{eqnarray}}
\newcommand{\eear}{\end{eqnarray}}
\newcommand{\eea}{\end{eqnarray}}
\def\hri#1#2{\href{http://arxiv.org/abs/#1}{[ArXiv:#1]#2}}
\def\hre#1#2{\href{http://arxiv.org/abs/#1/#2}{[ArXiv:#1/#2]}}

\newbox\pippobox

\def\parl{\parallel}
\def\dx{\delta X^\parl}

\def\II{\relax{\rm I\kern-.18em I}}

\def\e{\epsilon}
\def\l{\lambda}
\def\m{\mu}
\def\n{\nu}
\def\r{\rho}
\def\s{\sigma}
\def\pa{\partial}

\def\sp{\;\;\;,\;\;\;}

\def\a{\alpha}
\def\b{\beta}

\def\ls{\ell_s}

\def\parl{\parallel}

\title{Lorentz violation, Gravity, Dissipation and Holography}

\author{Elias Kiritsis$^{a,b}$\\
~\\
$^a$ \href{http://www.apc.univ-paris7.fr}{APC, Universit\'e Paris 7, Diderot},\\
 CNRS/IN2P3, CEA/IRFU, Obs. de Paris, Sorbonne Paris Cit\'e, \\
  B\^atiment Condorcet, F-75205, Paris Cedex 13, France (UMR du CNRS 7164)\\
~\\
$^b$ \href{http://hep.physics.uoc.gr}{Crete Center for Theoretical Physics},
Department of Physics, University of Crete,\\ PO Box 2208, 71003 Heraklion, Greece
}

\preprint{CCTP-2012-15 }

\abstract{We reconsider Lorentz Violation (LV) at the fundamental level. We show that  Lorentz Violation  is intimately connected with gravity and that
LV couplings in QFT must always be fields in a gravitational sector.
 Diffeomorphism invariance must be intact and the LV couplings transform as tensors under
coordinate/frame changes. Therefore searching for LV is one of the most sensitive ways of looking for new physics,  either new interactions or modifications of known ones.
Energy dissipation/Cerenkov radiation is shown to be a generic feature of LV in QFT.
A general computation is done in strongly coupled theories with gravity duals.
It is shown that in scale invariant regimes, the energy dissipation rate depends non-triviallly on two characteristic exponents,  the Lifshitz exponent and the hyperscaling violation exponent.
}


\begin{document}

\def\g{\gamma}
\def\go{\g_{00}}
\def\gi{\g_{ii}}

\maketitle 
\def\bu{$\bullet$}
\section{Introduction and historical perspective}

 Lorentz invariance is one of the most important pillars of modern day physics, experimentally unchallenged for more than a hundred years.
 It has been tested to unprecedent accuracy reaching { $10^{-21}$} in several contexts, \cite{tests} and down to { $10^{-29}$} recently in the neutron sector,
\cite{romalis}. Its reign coincided with the demise of ``aether".  Michelson and Morley
 first in a series of experiments and Einstein's special relativity in  1905
 eventually banished aether as a physical theory. Few physicists however realized
 that a new kind of ``aether" made it back into physics after 1915, with the general theory
 of relativity: the gravitational field.

Theorists rarely question Lorentz Invariance of physics and its universal grip.
Such investigations emerge from time to time. An interesting idea was put forward by H.
Nielsen and his collaborators, \cite{nielsen}, suggesting that Lorentz Invariance
is a low-energy phenomenon. Lorentz symmetry behaves as a conventional symmetry,
and suggests that vanishing Lorentz iolating couplings are always fixed points of
the renormalization group (RG). The nontrivial question is whether they are attractive fixed points.
 Nielsen and Ninomiya showed this in a special case, \cite{nn}.
However, the situation at large remains unclear.
Related ideas were applied to other low-energy symmetries, under the name  ``anti-grand-unification",
\cite{int}, but were not explored further than the original papers.

A different turn was taken in the late 80's.  Kostelecky and Samuel argued that in open string theory, unusual tachyon
vevs can trigger vevs of tensors and therefore  Lorentz Violation, (LV) \cite{ks}. Since then a lot has
been done in understanding the dynamics of tachyons in open strings
, \cite{sen}
and this understanding indicates that the original expectations in \cite{ks} do not bear out. However,
 the conjectured phenomenon is  dynamical Lorentz violation and so far there is no hint of it in quantum field theory, (QFT).
Kostelecky has followed-on in the analysis of the implications of LV, and several detailed parametrizations or LV in the Standard Model were
introduced and studied, \cite{kc}.

In the ninenties, several phenomenological Lorentz-violating dispersion
 relations have been contrasted to particle physics and astrophysical data starting with the works of Coleman and Glashow,
 \cite{Coleman} and later
\cite{aemns},  putting new stringent bounds on LV. Such dispersion relations were expected to
 be generated  among others from ``quantum gravity" or ``spacetime foam" models, but it is fair to say
that such expectations are still beyond the reach of understood techniques and frameworks.

In an unexpected turn, in the late nineties, and after the proposal of the
AdS/CFT correspondence, \cite{malda},  the
physics of probe D$_3$ branes was studied in a black D$_3$ brane background with
 the goal of computing the sYM effective potential at finite temperature and strong coupling, \cite{sym}.
It was noted that the effective speed of light on the brane was variable,
depended on the radial position of the brane
and could become much smaller than the bulk speed of light
(gravitational waves)\footnote{It was shown in full generality later in \cite{gh}
that the open string metric light-cone is always inside the closed string metric light-cone.},
 \cite{sym}. It was suggested that such a setup could be used as an alternative to inflation, by utilising the fact that
the bulk speed of light was larger than the speed of light on the brane, \cite{sym,freese}.

Brane models with a variable speed of light were subsequently constructed by stabilising branes
 in stable orbits around black branes, \cite{kir1,stephon,quevedo}.
 This idea can be brought to its logical conclusion by advocating a planetary brane universe, \cite{kir2}.
 It became also obvious that in orientifold constructions of the standard model (SM), not all SM ingredients emerge from the same
 D-brane. This affects certainly issues of unification, \cite{unif}, but it also implies that in a generic bulk gravitational background
 such realizations would provide SM particles with different light speeds. This would be particularly prominent for right-handed neutrinos as
 they would naturally emerge from the ``bulk"
 in such constructions\footnote{Here ``bulk" is used liberally to indicate branes that wrap the large spacetime dimensions. The fact that bulk neutrinos
  would be naturally light in the context of generic braneworlds was suggested earlier
  in \cite{neu}.}, \cite{unif,unif2}. Similar observations on LV were made on RS braneworlds in \cite{grosjean}.

 The idea that branes in nontrivial gravitational backgrounds have a varying speed of light,
  was brought to its natural conclusion in \cite{mirage}.
 There it has been argued that a probe brane in motion in a gravitational background
  (possibly generated by other branes) undergoes a cosmological evolution on its world volume that
 is not due to the brane energy density. It just reflects the presence of  bulk gravitational fields.
  This phenomenon has been termed ``mirage cosmology".
  A similar effect on RS braneworlds was described in \cite{kraus}, explaining
  ``dark radiation" in RS cosmology as a bulk effect.

The bulk geometries that were utilized to generate varying speed of
 light theories involved essentially black-brane geometries with
regular horizons. Such geometries have a non-trivial entropy associated
 to the regular horizon. Gubser, \cite{gubser}  has revisited the question of a varying speed of light
and  provided examples where the bulk geometry, has zero entropy.
Subsequently, many such  solutions were found as a byproduct of the
 holographic description of strongly coupled systems at finite density, \cite{taylor,cgkkm,gk}.

Lorentz violation in QFT was revisited in \cite{anselmi}. The key point was that an UV scale invariant QFT with Lifshitz scaling symmetry
has a different power counting structure, controlled by the dynamical exponent $z$. If $z$ is sufficiently larger than the Lorentz invariant value $z=1$, then
the standard hierarchy problem in a scalar sector may be vacuous.
Moreover in such cases, four fermion interactions may be renormalizable, and can be used as a substitute for electroweak symmetry breaking, \cite{wadia}.

Ho\v rava proposed  a similar idea in the gravitational sector, \cite{horava},
 using a UV theory of the gravitational field that breaks diffeomorphism invariance
but has $z=3$ Lifshitz scaling symmetry. This can provide a power-counting renormalizable theory
in four dimensions\footnote{Of course this does not make the theory renormalizable (UV complete). No calculation of the relevant
$\beta$-functions has been done so far to substantiate that the important interactions are UV marginally relevant.}.
It was pointed out, \cite{kk,mukohyama},  that a $z=3$ Lifshitz invariant UV theory of Ho\v rava gravity would contain the main ingredients to solve the
horizon and flatness problems of cosmology and generate scale-invariant perturbations without inflation.
The original Ho\v rava theory had several phenomenological problems in order to be considered as a valid substitute for standard gravitation, \cite{hlreview}.
A modified version, \cite{blas} was proposed that does not seem to have any obvious experimental conflicts.

The OPERA experiment in September 2011, \cite{opera}, has made the extraordinary
 claim, that neutrinos traveling from CERN to the GranSasso arrived earlier than
  expected suggesting that they moved with superluminal speeds. The news made the
   tour of the world prompting journalists to hail the demise of Einstein and
    physicists to ``remain sceptical to such an upheaval of special relativity". Since then,  several further tests have been made and the final experimental scrutiny indicated that
    the result was incorrect:
    A cable fault and a coarse timer were responsible for the unusual first result.

   The neutrino sector is more prone to the discovery of new physics as it is the most difficult to make measurements due to the weakness of interactions.
   Previous limits of Lorentz invariance include at a much lower energy (around 10 MeV), a stringent limit of ${v - c\over c} < 2\times   10^{-9}$ from
 the observation of (anti)neutrinos emitted by the SN1987A supernova, \cite{longo}.
  With a baseline analogous to that of OPERA,  but at lower neutrino energies (E peaking at 3 GeV with a tail extending above
100 GeV), the MINOS experiment reported a measurement of ${v-c\over c} = (5.1\pm 2.9)\times  10^{-5}$, \cite{minos}.
In the past, a high-energy ($E > 30$ GeV) and short baseline experiment, \cite{1},  was able to
test deviations down to ${v -c\over c} < 4\times   10^{-5}$.
The new (corrected) OPERA result is  an order of magnitude better at ${v -c\over c} <  10^{-6}$, \cite{opera2}.

In this paper we re-examine the issue of Lorentz violation in QFT, and we pair it with the fact that QFT is coupled to gravity. Our main conclusions are as follows:

\begin{itemize}

\item LV is intimately interlaced with gravity. LV couplings in QFT are fields in a gravitational sector. Diffeomorphism invariance is intact, and the LV couplings transform as tensors under
coordinate/frame changes.

\item Searching for LV is one of the most sensitive ways of looking for new physics: either new interactions or modifications of known ones.

\item Energy dissipation/Cerenkov radiation is a generic feature of LV.

\item A general computation can be done in strongly coupled theories with gravity duals. IN a scaling regime, the energy dissipation rate depends non-triviallly on two characteristic exponents,  the Lifshitz exponent and the hyperscaling violation exponent.

\end{itemize}

\section{Motivation and the main questions\label{que}}

In view of the fact that until now,  and for more than a hundred years, there is no experimental hint of the violation of Lorentz Invariance (LI), does it make sense to
theoretically explore Lorentz Violating Quantum Field Theories?
The answer is yes for at least four reasons:

\begin{itemize}

\item It is the main theoretical tool in contexts where LI is broken { by energy and charge densities}. The whole of condensed matter physics is in this
class.

\item The same argument is relevant in more exotic contexts e.g. quantum matter in the gravitational fields of a black hole and the strange phenomena (like Hawking radiation) that it entails.

\item It is important to understand the logistics of interplay between {quantum effects and symmetry breaking LV effects}. This might give models for the
spontaneous breaking, and guide experimental tests.

\item Last but most important, as it will argued in this paper, LV is one of the most efficient windows to new physics, both IR-sensitive and high-energy physics.

\end{itemize}

In view of the above we will pose  questions to be answered in the context of LV QFTs.

\begin{enumerate}

\item  {\it How is LV  compatible with diffeomorphism invariance and gravity as we know it?  Is gravity coupled to a LV QFT a consistent theory?}
This issue is important and will turn out to have far-reaching consequences. We will analyze it in section \ref{gravity}.

\item {\it If LI is violated how do we change coordinate systems?}. This question will also be addressed in section \ref{gravity} and will be shown to have a simple and predictable answer.

\item  {\it Can LI be spontaneously (dynamically) broken?} What is meant here is whether non-trivial vevs for tensor fields can be generated in the vacuum (in the absence of sources) in any LI QFT. Non-trivial vevs for vectors ($\ell=1$) occur naturally in condensed matter systems (nematic order), and cuprate superconductors have an $\ell=2$ condensate at low temperatures. However in such theories LI is broken by the finite charge density and energy.
    One may consider the question of U(1) vector vevs in a weakly-coupled quantum field theory. In such a context, a non-trivial vev will be generated if the quantum effective potential for the vector $A_{\m}$, has non-trivial minima. By Lorentz invariance, the effective potential has the form
    \be
    V_{eff}=F(A_{\m}A^{\m})
    \label{2.1}\ee
    where the function $F(x)$ has a global minimum at $x=x_{*}\not=0$.
    However, as is well known in QFT with $d>2$, a potential for a vector at weak coupling
    must be generated (foir gauge invariance purposes) via fields with minimal couplings. This is the only way a gauge potential appears without derivatives.
     Although there are several possibilities, we will assume here a charged scalar field\footnote{In general several charged fields could have a vev.} with a non-trivial vev.
    \be
    L_{A}=L_A(F)+G(D_{\m}\psi (D^{\m}\psi)^{\dagger},|\psi|^2)\sp D_{\m}\psi=(\partial_{\m}+iA_{\m})\psi
      \label{2.2}\ee
    where the function $G(x,y)$ will be considered a general function in two variables and $L_A$ is a LI function of the U(1) field strength.

   The general intuition is that if the function has a non-trivial minimum, triggering a vev for $A_{\m}$ in the vaccuum, this will imply that the kinetic term for the scalar will
   be ghostlike in that regime. Therefore, this will not be expected to happen in a healthy theory. This argument has loopholes and a more thorough investication of the question
   is required. We will not however pursue it further in this paper.

\item  {\it Is Lorentz invariance an accident of low energies?}  This is the question posed in \cite{nielsen,nn}. It was answered in the affirmative for non-covariant Maxwell theory.From the standard symmetry argument that Lorentz invariance does not seem to be broken during the RG flow of LI QFTs, it implies that scale invariant LI theories are fixed points of the RG flow. Moreover, at least at weak coupling,  it seems that the space of LI theories is closed under RG flow, (see however question 4).
    However, Lifshitz invariant theories can be also fixed points of the RG flow once LI is broken.
    It is well known that for several  Lifshitz invariant theories with $z>1$, there are relevant flows (via the operator that is related to kinetic terms by Lorentz transformations) that lead to LI RG fixed points.The opposite is also true in some cases: RG flows by LV operators of LI theories may lead to a Lifshitz IR fixed point.
    This however seems to involve tuning. To answer this question, a more global view of RG flows in LV QFTs is needed.

\item  {\it Under what conditions the breaking of LI is "natural" in the standard sense of Quantum field theory?} This is a question that can be in principle answered by standard methods and some steps in that direction have been made in  \cite{sh}.

\item {\it Does the effective speed of light $c_{eff}$ always decrease during RG flow?} This question is correlated with question 1. LI fixed point QFTs (CFTs) have finite values for the speed of light. Lifshitz QFTs with $z>1$ have an infinite speed of light. Therefore the rate of flow of the speed of light is determined by the nature of the UV and IR fixed point. We will address this question in more detail in section \ref{braneLV}.

\item  In the simplest cases at weak coupling LV is accompanied by energy dissipation via Cerenkov radiation.
{\it Is there a general connection between LV and energy dissipation?}
{\it Is there a difference between energy dissipation at strong coupling and weak coupling?}

\end{enumerate}

\section{Lorentz violation in QFT}

The first stop in this context is to parametrize LV in QFT.
We will take the simplest possible example, that of a real scalar field,  to indicate the issues. The generalization to more complicated QFTs is straightforward.

The most general action of this QFT is a linear combination of all local tensor operators, that we could  write as
\be
L=V_0(\phi)+{ V_1^{\mu}(\phi)}\partial_{\m}\phi+{ V_2^{\m\n}(\phi)}\partial_{\m}\phi\partial_{\nu}\phi+
 {  V_3^{\m\n\r}(\phi)}\partial_{\m}\phi\partial_{\nu}\phi \partial_{\rho}\phi+{\cal O}(\partial^2\phi) \cdots
\label{3.1}\ee
\be
V_n^{\m_1\m_2\cdots\m_n}(\phi)=\sum_{m=0}^{\infty} a_m^{\m_1\m_2\cdots\m_n}~\phi^m
\label{3.2}\ee
The constant numbers, $a_m^{\m_1\m_2\cdots\m_n}$ in (\ref{3.2}) should be thought of as generalized couplings.
All local operators appear in (\ref{3.1}). Total derivatives and operators that are related by the equations of motion are redundant and do not affect the bulk dynamics.

The general Lagrangian should be thought of as the bare Lagrangian of a UV QFT in the Wilsonian format.
In the most general case it can be replaced by a general CFT, a basis of the infinite number of its local (gauge invariant) operators, and a corresponding infinite
set of couplings. The operators can be classified by the irreducible representation of the $O(1,d-1)$ group they transform under.
The general LI action is composed of the (infinite) set of scalar operators of the theory. Adding couplings to the operators transforming non-trivially under O(1,d)
amounts to the most general LV action around the UV CFT.

Most of these operators are irrelevant and therefore not interesting for the IR physics. They may break  LI at higher energy scales.
 However there are some generic relevant or marginal operators that break Lorentz invariance.

 \begin{enumerate}

\item The (most) generic marginal operator is the stress tensor, $T_{\m\n}$. It is traceless (in a flat space-time) as this is the  stress tensor of the UV  CFT.
It is also marginal as it has canonical dimension $d$, and no anomalous dimension
as it is conserved in a translation invariant theory.\footnote{Translation invariance may be eventually broken if
the coupling constants become space-time-dependent. However, we may consider as the starting point the translational invariant UV CFT.
In particular, as there are always at least one point in coupling constant space where the theory is translationally invariant, this is sufficient to formulate
the expansion around this translationally invariant ground state, by considering arbitrary sources that break translation invariance.}

Turning on this operator always breaks Lorentz invariance, and corresponds to turning-on an energy-carrying source. A special case
corresponds to a thermal ensemble at fixed and finite temperature.
There can be no traceless spin-two tensors with scaling dimension smaller than $d$ by unitarity. All non-conserved spin-two tensors
 have scaling dimension larger than $d$ and are therefore irrelevant.
 If there is more than one spin-two tensor with dimension $d$,  the CFT is a direct product of several CFTs, \cite{multi}.

\item Another non-trivial tensor operator can be a conserved vector ( i.e. a conserved current). Such vectors are associated with global
symmetries that are intact and have scaling dimension $d-1$, modulo exceptions that are discussed below. Turning them on breaks LI, and generates a state with finite charge.
Any other non-conserved vector has dimension larger than $d-1$, and as long as it remains smaller than $d$ it can be a relevant LV operator perturbation.

\item Finally a third possibility is a two-index antisymmetric tensor, $B_{\m\n}$. Unitarity constraints its scaling dimension in a CFT to be $\Delta\geq 2$.
Such operators can be decomposed into selfdual and anti-selfdual parts, $B_{\m\n}^{\pm}$,  transforming as the (1,0) and (0,1) representations of the Lorentz group.

A trivial case corresponds to total derivatives that are not relevant here: for any vector operator $T_{\m}$ one can write $\pa_{\m}T_{\n}-\pa_{\n}T_{\m}$.
This operator does not affect the dynamics if added to the action with a constant source.
If the source $J^{\m\n}$ is not constant then the coupling can be written as
\be
\delta S=\int d^4x ~J^{\m\n} (\pa_{\m}T_{\n}-\pa_{\n}T_{\m})\simeq -2\int d^4x ~(\pa_{\m}J^{\m\n}) ~T_{\n}
\ee
where the $\simeq$ sign implies an integration by parts and dropping the surface terms. Therefore in the general source functional such a term can be absorbed to a redefinition of the source of the vector operator $T_{\m}$.

It follows from equation (5.15) of \cite{petkou} that a non-trivial operator $B_{\m\n}$ with the minimal scaling dimension $\Delta=2$ is conserved $\pa^{\m}B_{\m\n}=0$. 
For any dimension $\Delta>2$, the vectors $J_{\m}=\partial^{\n}B_{\n\m}$, and $\tilde J_{\m}={1\over 2}{{\e_{\m}}^{\n\r}}_{\s}\pa^{\s}B_{\n\r}$ of dimension $\Delta+1$ are automatically conserved $\partial^{\m}J_{\m}=\partial^{\m}\tilde J_{\m}=0$ due to the antisymmetry of $B_{\m\n}$. Therefore a single $B_{\m\n}$ can be decomposed as $B_{\m\n}=B^{+}_{\m\n}+B^-_{\m\n}$
and the two conserved currents can be written as $J^{\pm}_{\m}=\partial^{\n}B^{\pm}_{\n\m}$. It is important to note that the conserved currents have non-canonical dimension $\Delta+1>3$.
It is plausible that such non-canonical conserved currents exist only in free CFTs.  

There are not many examples of such operators that are relevant or marginal and all we know come from free CFTs.
The lowest dimension example appears in the free field CFT of a massless free photon. The gauge invariant operator is the 
abelian field strength $F_{\m\n}$, of dimension $\Delta=2$. Although this is of the derivative form discussed above, the relevant vector operator $A_{\m}$ is not gauge invariant, therefore the operator $F_{\m\n}$ is non-trivial.
The conservation equation $\partial^{\m} F_{\m\n}=0$ implies that the photon is free. 

In the free CFT of massless fermions we have the dipole operator $B_{\m\n}=\overline\psi \gamma^{\m\n}\psi$ with dimension 3. It is therefore relevant.
The associated conserved vectors of dimension 4, are $\overline\psi\overleftrightarrow{\pa}\psi$ and $\overline\psi\overleftrightarrow{\pa}\gamma^5\psi$.
In theories with a holographic dual typically this operator is associated to a massive string mode and is therefore irrelevant.

In the presence of at least two scalars
we may write $B_{\m\n}=\pa_{\m}\phi_1\pa_{\n}\phi_2-\pa_{\n}\phi_1\pa_{\m}\phi_2$. This is an operator of dimension 4 in the free field limit and may become
marginally  relevant. The associated conserved currents of dimension 5 are $\pa^{\m}\phi_1\pa_{\m}\pa_{\n}\phi_2-\pa_{\m}\pa_{\n}\phi_1\pa^{\m}\phi_2$

Note that in all free field cases, the presence of an non-trivial two-form operators is correlated with the presence of a continuous U(1) symmetry.
 
In both this case and the fermionic one, once the free field acquire a mass, the conserved higher dimension currents acquire always an admixture of the 
canonical U(1) current. In the Maxwell case, the presence of a mass for the photon breaks the gauge symmetry. In this case, the gauge field $A_{\m}$ is a conserved current, $\partial_{\m}A^{\m}=0$, and is given by the divergence of the two-form, $\pa^{\m}F_{\m\n}\sim A_{\n}$.

\end{enumerate}

It is interesting to note that the (space-dependent) coupling constants to such operators have concrete and known interpretations:

 \begin{itemize}

\item The coupling constant of the conserved stress tensor $T_{\m\n}$ is a metric. This metric gauges the translations of the QFT.

\item The coupling constant of a conserved current of canonical dimension is a gauge field. It gauges the associated global symmetry.

\item The coupling constant of a two form operator $B_{\m\n}$ is a two form gauge field.

\item The coupling constant of a non-canonical current $\pa^{\m}B_{\m\n}$ is a gauge field coupled magnetically (via dipole coupling) to the QFT.

\end{itemize}

Due to unitarity, this exhausts all possible relevant or marginal LV bosonic operators in a CFT. All other relevant bosonic operators must be Lorentz scalars.

On the other hand there can be fermionic relevant operators that can be used to break Lorentz invariance via fermionic coupling constants.
A generic spin-1/2 fermion operators has dimension that is larger or equal to ${d-1\over 2}$ and can be relevant. A conserved spin-3/2 operator
has necessarily dimension $d-{1\over 2}$ and is relevant. Such operators exist in QFTs with unbroken supersymmetry.
If supersymmetry is broken such operators acquire anomalous dimensions but may remain relevant.
This exhausts all nontrivial LV relevant operators in a CFT. It is fair to say though that fermionic perturbations seems unusual and we know of no context where they
can appear in a  physical effective action.

Therefore at low energies LV is due to effective couplings that turn on energy,  (approximately conserved) charges and generalized electromagnetic fields.

Going back to our model example theory in (\ref{3.1}), for generic values of the couplings LI is broken.
There are however special values that retain LI.
For example when   $a^{\m}_m=0$, $a^{\m\n}_m\sim \eta^{\m\n}$, $a^{\m\n\rho}_m=0$, $\cdots$, the theory is LI.
When $a_m^{\m_1\m_2\cdots\m_n}$ become functions of the space-time point, then translation invariance is broken.
If the theory is coupled to a non-trivial metric then we would expect that
\be
a_m^{\m\n}=c_m\sqrt{-g}~g^{\m\n}
\label{3.3}\ee
with $c_m$ constants.
This indeed breaks LI , and we will call it {\it ``environmental breaking"}. According to GR, a non-trivial metric is the outcome of a non-trivial energy density, and is therefore non-surprising
that it breaks LI. This example bring to forth an important qualification when it comes to identify LV effects: they should be separated from the ``obvious" ones.

The {\it obvious} effects contain:
\begin{itemize}
\item Energy and charge distributions  in the theory in question.

\item Other long range fields generated by such distributions (like electromagnetic fields).

\item Energy and charge distributions in other/uncknown sectors of the theory, whose effect is to generate a non-trivial gravitational field.

\end{itemize}

This should be distinct from what we will call {\it dynamical} LI breaking. This is defined as follows: We consider a standard QFT with fields of spin up to one.\footnote{
One may attempt to extend the definition to a larger class of quantum theories that may contain theories
with spin two fields, as in the Ho\v rava paradigm, \cite{horava}. All such theories however in order to make sense as quantum theories must be defined by a non-trivial
 Lifshitz asymptotics in the UV. Therefore LI is broken explicitly and issue of dynamical breaking looses its meaning.}
If there are non-trivial {\it vacuum} expectation values that transform non-trivially under $O(1,d)$ then we call this {\it dynamical } LV.
To our knowledge there is no such example known neither a no-go theorem. It is unlikely it can occur at weak coupling.

As usual in QFT we may consider couplings of QFT operators to arbitrary
sources\footnote{If a gauge symmetry is present in the QFT, then only gauge invariant local operators are considered.}
 as in (\ref{3.1}), (\ref{3.2}) and consider the generating functional of correlation functions,
\cite{tom},
\be
e^{-W(a_i)}=\int {\cal D}\phi e^{-\int d^{d+1}x \sum_i a_i(x) O_i(x)}
\label{3.4}\ee
where $O_i$ labels the infinite number of local QFT operators, $\phi$ denotes the set of quantum fields and $a_i$ are the associated sources.
The functional of the sources $W$ is a complicated functional that in the absence of a global mass gap is non-local.
When treating the sources as infinitesimal, $W$ is the main tool in the calculation of the (connected) correlation functions,
\be
\langle \prod_{j=1}^{N}O_j(x_j)\rangle_c= \prod_{j=1}^{N}{\delta \over \delta{a_j(x_j)}}~W(a_i)\Big |_{a_i=0}
\label{3.5}\ee
It is a less known fact that the global symmetries, like translation invariance and particle conservation of QFT become local symmetries for W.
This is easy to show at a linearized level. Consider a QFT which is translationally invariant as well as invariant under a U(1) global symmetry.
This translates into the transformation properties of the QFT action under infinitesimal transformations as
\be
S'=S+\int d^{d+1}x \left[2T_{\m\n}(\partial^{\m}\e^{\n})+ J_\mu(\partial^{\m}\e)+{\cal O}(\e^2)\right]
\label{3.6}  \ee
where $T_{\m\n}$ is the symmetric conserved stress tensor\footnote{There are several theories
where there can be subtleties with the stress tensor but the argument made here is eventually correct.},
$J_{\m}$ is the conserved U(1) current, $\e^{\m}(x)$ is a local infinitesimal translation parameter,
 ($x'^{\m}=x^{\m}+\e^{\m}$) and $\e(x)$ is a local infinitesimal U(1) transformation parameter.
When $\e^{\m},\e$ are constants, the action of the QFT is invariant and the associated transformations symmetries that imply that
\be
\partial^{\m}T_{\m\n}=0\sp \partial^{\m}J_{\m}=0
\label{3.7}\ee
Consider now the action $S$ coupled to sources
\be
S(A_{\m},h_{\m\n})\equiv S+\int d^{d+1}x\left[h^{\m\n}T_{\m\n}+A^{\m}J_{\m}\right]
\label{3.8}\ee
In view of (\ref{3.6}) the action with sources (\ref{3.8}) is invariant under the local transformations
\be
S(A'_{\m},h'_{\m\n})=S(A_{\m},h_{\m\n})+{\cal O}(\e^2,\e h,\e T)
\label{3.9}\ee
with
\be
A'_{\m}=A_{\m}-\pa_{\m}\e\sp h'_{\m\n}=h_{\m\n}-\partial_{\m}\e_{\n}-\partial_{\n}\e_{\n}
\label{3.10}\ee
The linearized local invariance, can be promoted order by order to a bona-fide diffeomorphism and U(1) gauge invariance by the Noether method.
In the end, the source action becomes locally diffeomorphism and gauge invariant.
The final part of the argument, involves the construction of a locally invariant path integral integration measure in (\ref{3.4}).
When this is possible then the full generating functional of the sources, is invariant under the nonlinear transformations.
\be
W(g'_{\m\n},A'_{\m},\cdots)=W(g_{\m\n},A_{\m},\cdots) \sp g'_{\m\n}=g_{\r\s}{dx^{\r}\over dx'^{\m}}{dx^{\s}\over dx'^{\n}}\sp A'_{\m}=A_{\m}-\pa_{\m}\e
\label{3.11}\ee
Obviously, other sources must transform appropriately under the diffeomorphisms and local gauge symmetries.
The above applies to all abelian and non-abelian global continuous symmetries.
When the global symmetries are anomalous there are modifications.
There are two types of anomalies that are relevant in this discussion:

\begin{itemize}
\item Non-conservation of global currents in the presence of external sources. This is the case of the chiral symmetries  of a free fermionic theory
as a simple example. In this case, under a U(1) gauge transformation
\be
\delta W(g_{\m\n},A_{\m},\cdots)={\rm Anomaly}(g_{\m\n},A_{\m},\cdots,\e)
\label{3.12}\ee
where the form of the anomaly depends on the symmetry and the dimension. In four dimensions, and for a U(1) global symmetry we have
\be
 {\rm Anomaly}(g_{\m\n},A_{\m},\cdots,\e)=c\int d^4x~\epsilon^{\m\n\r\s}~F_{\m\n}(A)F_{\r\s}(A)~\e+{\cal O}(\e^2)
 \label{3.13}\ee
 where $c$ is a constant.

\item Non-conservation of  global currents due to gauge interactions and instantons. An example in this case is the $U(1)_A$ in QCD.
In this case, there are other sources that involved in the anomalous gauge transformations. For example in the case of the $U(1)_A$ in QCD
it is the source $a$ of the topological term (similar to an axion field) as well as the phase of the source of the quark mass operator $\theta$.
In this case, the effective action is invariant if together with the U(1) gauge transformation, there is an appropriate shift in $a$ and $\theta$ in the effective action, (see \cite{axion} for an associated holographic realization).
\end{itemize}

An important question is: can we always promote a linearized gauge invariance to a full non-linear gauge gauge symmetry by the Noether method.
The answer seems to be in affirmative, \cite{marc}, the only obstructions in this context being anomalies that were discussed above.
Sometimes, like in the case of diffemorphism invariance, there is an infinite number of steps required, but the procedure is convergent. It may also
be possible by a clever choice of variables to be made to converge much faster, \cite{deser}. The relevant symmetry identities may be needing  redefinitions in order to be properly realized,
\cite{porrati}.

 We therefore conclude that in the effective functional for the sources, upon a proper non-linear redefinition of the sources, all global symmetries of the QFT  (even spontaneously broken ones) are realized as {\it local symmetries} on the sources, including diffeomorphism invariance reflecting the translational symmetry of the original CFT.
 A related functional that is of importance in QFT is the effective action, the generating functional of 1PI diagrams in perturbation theory, defined as the Legendre transform of $W$ with respect to the sources
 \be
 S_{eff}(b_i)=\int d^{d+1}x~ \left[\sum_i a_i b_i\right]-W(a_i)\sp b_i(x)\equiv {\delta W\over \delta a_i(x)}\sp {\delta S_{eff}\over \delta b_i(x)}=a_i(x)
\label{3.14} \ee
$b_i$ is the (quantum) expectation value of the QFT in the presence of sources.  In particular the vacuum expectation values of the QFT in the absence of sources are obtained by solving the classical equations
\be
  {\delta S_{eff}\over \delta b_i(x)}=0
\label{3.15}  \ee
The source functional as well as the effective action may depend a priori from an infinite number of sources, dual to all possible local operators of the QFT.
If we fix a number of sources, then the dynamics for the rest of the expectation values is obtained by partially solving some of the equations
${\delta S_{eff}\over \delta b_i(x)}=a_i(x)$ for the appropriate sources.

The discussion above indicates that a parametrization of Lorentz violation in QFT is intimately related to the effective source functional $W$, for all sources.
The non-trivial ones from the point of view of LI are the ones that transform non-trivially under Lorentz transformations.
The effective source functional for a QFT,  $W$,  is a remarkable functional that contains only local symmetries.
 All global symmetries of the associated CFT have transmuted to local symmetries {\it always} including diffeomorhism invariance\footnote{As for any QFT there is a point in coupling constant space where it is translationally invariant. It would be interesting to know whether there are counterexamples to this expectation.}.

At this stage we observe that several of the natural Lorentz breaking (but also Lorentz preserving sources) are typically dynamical fields in a gravitational sector.
The case for the metric (source for the stress tensor) is obvious. More sources  however may play the same role. The source coupled to the topological (instanton) density is an axion field
not very different from the Peccei-Quinn axion.
The (scalar) couplings multiplying fermion mass terms, Yukawa couplings, Higgs potential terms can be scalars in the gravitational sector.
Finally, sources coupled to conserved currents may be graviphotons.

This scheme of affairs is not surprising as it is the generic picture  we obtain in string theory.
In particular in the case of open string theory and orientifolds, there is a natural separation between
what we could call QFT (open string sector with fields at the lowest level
having spin up to one), and the gravitational sector generated by the closed strings.

There seems to be however, an important difference between the source functional $W$ of a QFT and
the associated effective action of a closed+open string theory, that contains both QFT
 (under the guise of the open sector) and the gravitational sector from closed strings serving also as sources for the open sector.
In the first case the sources are not dynamical while in the second (string theory) case they are.

The case of the AdS/CFT correspondence is rather suggestive in this respect.
The correspondence states in its generalized form that the source functional $W(a_i(x))$ for a large-N QFT
is the same as the S-matrix, $S(a_i(x))$,  of a dual string theory in a concrete semiclassical background that is
asymptotically AdS, with an AdS boundary (conformally) isomorphic to the space-time of the QFT.

There are several subtleties in this correspondence.
First, $a_i$ are sources for single trace operators only. They are in one
 to one correspondence to single-particle string states.
Multiple trace operators correspond to multi-particle string states.
 As the string coupling is $g_s\sim {1\over N}$ , at weak string coupling (large N)
single particle states are clearly distinguishable from multiparticle states
 as they are nearly stable. However at finite and small N, this distinction is not at all clear.
In this case we expect non-locality to come in and in the string theory side,
it it fueled by string loop effects.

An important point is that the string S-matrix is a function of the boundary sources, that are in themselves
 non-dynamical (as they are also on the QFT side).
It is the bulk string fields that are dynamical, but the bulk dynamics in emergent in the AdS/CFT correspondence.
Both the holographic directions as well as other possible internal dimensions (like the $S^5$ is the most symmetric
example) appear from the large-N eigenvalue distributions of QFT quantum fields, \cite{MV}.

In the case of open string theory, whose basic zero mode spectrum is that of a QFT, containing fields of spin up to one, the coupling to sources
corresponds to the coupling to the closed string sector. In this case it is known that the UV divergences of the one-loop open string amplitude reproduce
poles that are interpreted as the closed string massless poles. This is the reason open string theory without a UV cutof is consistent only if coupled to closed string theory.

Turning this argument around, the quantum physics and consistency of the open string theory requires that closed strings are dynamical.
Moreover, the closed string equations can be interpreted as RG flow of the opens string theory. Preliminary suggestions to that effect appeared in \cite{BA,dfkz}.
The AdS/CFT correspodence added a strong incentive to take this seriously.
Indeed, in \cite{KV} it was shown that at weak coupling the open string diagrams satisfy RG equations that are similar to the closed string equations of motion.
Although the arguments were made at finite $\alpha'$, if the AdS/CFT cojecture is correct they should be valid in the QFT ($\alpha'\to 0$) limit.
    If this is the case then we may state generally that:
\vskip 0.3cm
\centerline{\it  Gravity/string theory is the Renormalization group dynamics of the sources of QFT.}

\vskip 0.3cm
It is an interesting open problem to show that indeed such a statement is correct in general.

\section{The coupling to the gravitational sector\label{gravity}}

The question we would like to address here is how do we couple a LV QFT to gravity\footnote{Couplings of Lorentz violating theories to gravity were considered before, \cite{ko1}.
They have not reached however the conclusions we reach here, namely that the LV couplings are fields in the generalized gravitational sector.}.
To address this we should remember that gravity is making translations and Lorentz transformations local (gauge)  symmetries.
Therefore it would seem that it is impossible to couple a LV QFT to gravity.
The answer however lies in similar situations with standard gauge theories as well as local supersymmetry in the context of supergravity and string theory.
There are two options (a) either restore the symmetry in the QFT using extra degrees of freedom or (b) break the gauge symmetry.
Unfortunately, in more than two dimensions option (b) is not feasible as breaking explicitly the gauge symmetry provides a sick quantum theory.
Therefore the only leftover option is to restore the symmetry.

The restoration can be done with the standard Stuckelberg algorithm, and we will show it in a simple U(1) example.
We start with a ``matter Lagrangian" for a complex scalar that is not invariant under the standard U(1) global symmetry,
\be
L=|\partial \psi|^2+Re[g\psi^2]\sp \psi\to \psi ~e^{i\e}
\label{4.1}\ee
The U(1) non-invariance could be restored if the coupling constant $g$ transforms appropriately.
To couple this theory to a U(1) gauge theory of a photon, we must introduce a scalar degree of freedom $a$ that will restore the U(1) invariance
of the scalar potential. Moreover there must be an appropriate term that locks $a$ to the transformation of the gauge field, and this
 is the standard Stuckelberg term generating a photon mass.
 The gauge invariant action is then
\be
L'= | D_{\m} \psi|^2+Re[g\psi^2 e^{-2iqa}]-{1\over 4e^2}F_{\m\n}^2-{m^2\over 2}(\pa_{\m} a+A_{\m})^2\sp D_{\m}\psi\equiv (\partial_{\m}+iq A_{\m})\psi
\label{4.2}
\ee
which is gauge invariant under
\be
A_{\m}\to A_{\m}-\pa_{\m}\e\sp a\to a+\e\sp \psi\to \psi ~e^{iq\e}\;.
 \label{4.3}
 \ee
The upshot of this construction is that the symmetry violating couplings become fields that transform so that the associated invariance is restored.
Translating this to our problem indicates that the only consistent way a LV theory can be coupled to gravity is if the LV couplings are promoted to
fields in the gravitational sector that transform according to the standard rules of coordinate invariance.

For example the LV vector coupling $L_1=\int d^{d+1}x~b^{\m}\partial_{\m} \phi(x)$ must be promoted to $L_1'=\int d^{d+1}x~\sqrt{g}g^{\m\n}A_{\m}(x)\partial_{\n} \phi(x)$
with the new vector field transforming as vector under general coordinate transformations. In a Minkowski background $A_{\m}=b_{\m}$ provides the original LV theory.

The general picture is clear. The source effective action provides a coordinate-invariant functional if all the sources transform as appropriate tensors under
coordinate transformations.
This implies the following general statement:\\
{\it All sources of Lorentz violation in QFT are environmental and can be associated with tensor fields in a ``gravitational sector" having non-trivial expectation values}.
Although this statement looks natural if cast in the light of the above discussion, it is not always so.
For example Lorentz violation emerging from non-commutativity of coordinates, may a priori look different. However upon further thought it transpires that
it can be equivalently generated by backgrounds fields, as was first shown in open string theory in \cite{sw} and more generally in recent works, \cite{lust}.

Earlier we distinguished two types of LV namely dynamical and environemental one. Does dynamical LV, if possible at all, adapts to a coupling to gravity as described above.
The answer is an obvious yes. In the LV vaccum, expanding around  the non-trivial vevs, the theory still looks like the case of the environmental LV, the only difference being that the LV couplings now are due to the QFT itself.
Therefore there is no difference when we consider the coupling to gravity from what we argued earlier.

At this stage we can answer the question that has created a lot of confusions in the literature: {\it If LI is broken, how are we supposed to change coordinate systems?}
The answer implied by our previous conclusion is that:\\
{\it General coordinate invariance remains intact, and we change general coordinate frames as Einstein postulated in 1915.
 We must however transform the LV couplings as tensors
on the way.}

A certain puzzle however looms on the way: if all LV theories compatible with theory are general coordinate
invariant then where does Ho\v rava-Lifhsitz gravity stand? We will discuss this issue in the next subsection.

Before moving on we should comment on the term ``gravitational sector" that has been liberally used so far in the discussion.
Is the gravitational sector a well defined concept? The short answer is no, appart from the fact that it contains at least a (massless) graviton.
The closest example of a more or less clean separation of a gravitational sector exists in type-II orientifolds in string theory.
There,  the open string sector that is comprised of D-branes and generates ``matter" (The SM is typically embedded in this part, \cite{kir2}) is similar to the QFT sector.
There is also the closed string sector that contains among others gravity.
However, even in this best of cases, we know that non-perturbative dualities sometimes mix these two sectors. For example the standard Heterotic-type IIA duality
as well as Heterotic type I duality maps both open and closed sectors to a closed sector, \cite{arge}.

\subsection{The Ho\v rava-Lifshitz paradigm and coordinate invariance}

The Ho\v rava construction on non-relativistic gravity rests on a UV theory with Lifshitz scale invariance.
This is a LV scale invariance where time and space scale differently,
\be
t\to \l^{z}~t\sp x^i\to \l ~x^i
 \label{4.4}\ee
The {\it ``dynamical exponent"} $z$ can be  different from the LI Case ($z=1$). In all known well-behaved examples, it is always $z\geq 1$.
A simple free-field realization  with $z=n\geq 1$ is the following
\be
L=\dot \phi^2-\phi~\square^{n}\phi+\cdots\sp {z=n}
 \label{4.5}\ee
 The $\square$ is the Laplacian in the spatial directions only. It is for this reason that the  theory does not contain ghosts although it has higher derivatives in the quadratic part.

The $z=2$ case is realizable experimentally in condensed matter first in antiferromagnets, \cite{lif1} and more recently in double-layered graphene, \cite{lif2}.
Such theories can be obtained by tuning the coefficient of the $\phi\square\phi$ term to zero. As this coefficient is proportional to ${1\over c^2}$ this implies that
in such scale invariant theories with $z>1$, the speed of propagation of the $\phi$-waves is infinite, or equivalently that there is no $\phi$-cone.
Such theories  could be bona-fide UV fixed points that will define non-trivial RG flows.

Lifshitz scaling theories have improved UV behavior as can be seen from the (free) propagator of the theory in (\ref{4.5})
\be
\langle \phi(t,\vec x) \phi(0,\vec 0)\rangle~~\sim~~i\int {dE d^dp\over (2\pi)^{d+1}}  ~{i~e^{iEt+i\vec p\cdot \vec x}\over { E^2-(\vec p)^{2z}}}
 \label{4.6}\ee
 Indeed, as $z>1$ the propagator has a faster fall-off at large $\vec p$ than the $z=1$ propagator. This implies a different power counting structure in QFTs that flow from Lifshitz fixed points.  Simple dimensional analysis can make this precise. For the free theory in (\ref{4.5}) we obtain
\be
S=\int dt ~ d^{d}x\left[\dot \phi^2-\phi~\square^{z}\phi+\cdots\right]\sp [\phi]={ {d-z\over 2}}
 \label{4.7}\ee
Therefore in 3+1 dimensions, if {$z=3$}, all $\phi^n$  terms in the action are renormalizable terms as $[\phi]=0$. If $z=2$, terms up to $\phi^{10}$ are renormalizable.
 For fermions, {$[\psi]={d-1\over 2}$}, and if {$z=2$}, the Nambu-Jona-Lasinio type of interactions
\be
S_{NJL}\sim g_{ijkl}\int ~~\bar\psi^i\bar\psi^j\psi^k\psi^l
 \label{4.8}
 \ee
 are renormalizable (marginal).
Where it not for the LV in this case one could break EW symmetry without fundamental scalars therefore avoiding the  hierarchy problem.

In Lifshitz theories the term in the action that turns-on a finite $\phi$-speed, namely $\phi\square\phi$ is always relevant,
and turning it on in the UV it is expected to lead the theory to an IR fixed point that is LI.

Ho\v rava's suggestion is that a $z=3$ Lifshitz theory theory for the gravitational field, would be power coupling renormalizable in 3+1 dimensions despite the broken diffeomorphism invariance, \cite{horava}.
To define the theory, the (3+1)-dimensional metric is ADM decomposed as
\be
ds^2=-{ N}^2
~dt^2+{ g_{ij}}(dx^i+{ N^i}dt)(dx^j+{ N^j}dt)
 \label{4.9}\ee
The kinetic terms are given in terms of the extrinsic curvature as
\be K_{ij}={1\over 2N}(\dot
g_{ij}-\nabla_{i}N_j-\nabla_jN_i)
 \label{4.10}\ee
\be
S_K={2\over \kappa^2}\int dtd^3x
\sqrt{g}N~\left(K_{ij}K^{ij}-{ \l} K^2\right)
 \label{4.11}\ee
The rest of the action does not contain time derivatives on $g_{ij}$.
To determine what kind of terms are renormalizable in the rest of the action we must specify the scaling of the metric components
\be
t\to \l^{ 3} ~t\sp x^i\to \l~x^i\sp
[N]=0\sp [g_{ij}]=0\sp [N_i]=2
 \label{4.12}\ee
Then  the ``potential" is a function of $g_{ij}$ and its spatial derivatives.
\be
V=\int dt d^3x~\sqrt{g} N~V(g_{ij})
 \label{4.13}\ee
For renormalizability it should contain up to six derivatives.  The six-derivative
  terms are classically Lifshitz-scale invariant.
Terms with a lower number of derivatives are "relevant".
The six derivative terms that contribute to the graviton propagator are
\be
\nabla_{i}R_{jk}\nabla^{i}R^{jk}\sp \nabla_{i}R_{jk}\nabla^{j}R^{ik}\sp R\square R\sp R_{ij}\square R^{ij}
 \label{4.14}\ee
while the rest are
\be
R^3\sp RR_{ij}R^{ij}\sp R_{ij}{R^i}_{k}R^{jk}
 \label{4.15}\ee
and provide scale-invariant interactions. There are further terms that involve the lapse $N$ and which were introduced in \cite{blas}.
The (local) invariance of the theory is
\be
t\to h^0(t)\sp x^i\to h^i (t,x^j)
 \label{4.16}\ee
and falls short of the full diffeomorphism invariance.
Renormalizability has not yet been tested at the loop level.

According to our previous conclusions, there seems to be no place for Ho\v rava-Lifshitz gravity coupled to LV QFT.
There is however a loophole in our previous argument. We have defined the QFT as a flow of a UV CFT which was LI. All its perturbing operators
were classified according to the $O(2,d+1)$ invariance, and this leads to general coordinate invariance.

The situation would have been different if the UV CFT used to define the Lorentz violating QFT was a CFT with non-trivial Lifshitz symmetry.
In that case the infinite number of operators are classified by the Lifshitz symmetry that is substantially smaller than $O(2,d+1)$.
This symmetry contains apart from scale transformations, space rotations.
In such a case the natural gravitational theory that the QFT can couple to is a (generalized) Ho\v rava-Lifshitz gravity.
A quick way to see this is to check that the renormalization countertrms of holographic Lifshitz theories generate
Ho\v rava-Lifshitz gravity, \cite{griffin}.

However the two approaches are related by an (infinite) reorganization of the operator content, and upon such a reorganization they should agree.
Indeed, the Ho\v rava-Lifshitz theory can be written as a coordinate invariant theory using the standard Stuckelberg method, by promoting the
general reparametrizations of time to a scalar field, \cite{blas2}. Moreover the modified theory proposed by \cite{blas} is related to a subsector of the Einstein-Aether theory,
\cite{jaco,kir-n} which is a generally covariant theory.
This concludes the arguments and welcomes Ho\v rava-Lifshitz gravity in the zoo of coordinate-invariant gravitational theories.

\subsection{Conclusion}

 We have shown in this section, under very weak assumptions, namely the necessity of preserving  coordinate invariance, that {\it any Lorentz violation in
QFT is associated with background fields of a generalized gravitational sector.}
This conclusion that indicates that all LV is environmental (spontaneous breaking, if it exists in QFT,  excluded), turns the issues of LI upside down.
What we have shown is that
$$
Lorentz~~ violation ~~~\simeq~~~~ {`` fifth~~ forces}"
$$
In particular,  an unknown new force, mediated by a scalar, vector,  spin-two tensor etc may  couple to (some of the)  SM particles.
SM particle coglomerations will generate a classical non-trivial background for the new field, that will generically break LI.
The field would be long range if the new force has a nearly massless carrier while it will be short range otherwise.
Detecting the force will be more difficult in the second case.

The  initial OPERA claim \cite{opera} sparked a flury of models attempting to explain the experimental result.
From the consistent models proposed, most attempted to accommodate the result by parametrizing appropriately the neutrino couplings in the SM.
The two extra alternatives proposed included a new massive graviton interaction, \cite{dvali} and a new scalar interaction \cite{kehagias}.

In view of the above, can Lorentz violation originate from a known force? The answer is in the affirmative if the dynamics of that force turns out to be different from what we think,  in a regime of energies/distances.
For example,  a potential modification of the gravitational interaction at long distances could be potentially interpreted as LV in concrete experiments. The reason
is that the  interpretation of observed LV is dependent on subtracting the effects of known sources like the gravitational source.
Therefore incomplete knowledge of the gravitational force
would result in incomplete subtraction and therefore a ``signal".

In all the cases above LV is related to new physics, most importantly IR-sensitive new physics.
{\it Because of the astounding strigent limits, \cite{sensi}, experimental searches for LV, are probably the most sensitive probes of IR-sensitive new physics.}

\section{Lorentz Violation on Branes\label{braneLV}}

In this section we will review how the apparent speed of light on branes
embedded in a higher dimensional bulk may be made variable due to bulk gravitational
fields, \cite{sym}. The bulk gravitational fields via their coupling to the brane
theory will provide in this case the source of LV. This is a simple example of LV that
will serve as a laboratory, in order to connect this via holography to strong coupling physics. It will allow us to find
strongly-coupled analogues of energy dissipation further on.

We will consider a generic black 3-brane metric in a 5 dimensional bulk
\be
ds^2=e^{2A(r)}\left[-f(r)dt^2+{dr^2\over f(r)}+dx^idx^i\right]
 \label{5.1}\ee
Here $t,x^i$ are 3+1 coordinates while $r$ will be a fifth radial coordinate.
$e^{A}$ is the scale factor of the metric and  $f(r)$ is the ``blackness" function.
The space will have a boundary at $r=r_{\infty}$ where the scale factor diverges $e^A(r_\infty)=\infty$
and a horizon at $r=r_h$ where as usual $f(r_h)=0$.
The blackness function monotonically decreases from a constant value at the boundary to zero at the horizon\footnote{It can be shown, \cite{gubser-n} that
$f$ cannot have a regular extremum unless the null energy condition is violated. Here the extrema happen at the boundary and at the horizon.}
\be
0\leq f\leq c_{UV}^2\;.
 \label{5.2}\ee
 The boundary constant value $c_{UV}$ is identified with the speed of gravity in the bulk as well as
with the UV speed of light in the holographic dual.

We now consider embedding a 3-brane at $r=r_*$. Typically,  this embedding is not  stable. It can be stabilized by giving angular momentum in
extra dimensions, \cite{kir1,stephon,quevedo}. It may also be stabilized by including an extra vector  interaction (like the RR interaction in D-branes).
We now consider the effective theory on the 3-brane. We will assume for simplicity that there is a photon on the 3-brane with a standard Maxwell kinetic term.
Similar statements will apply to other localized particles and interactions.
\be
S_{brane}= -{1\over 4}\int d^4x~\sqrt{-\det(g)} ~F_{\m\n}^2+\cdots
 \label{5.3}\ee
with the induced metric
\be
d\hat s^2=\hat g_{ab}d\xi^a d\xi^b=G_{\m\n}~{\partial X^{\m}\over \partial\xi^a} ~{\partial X^{\n}\over \partial\xi^b}d\xi^a d\xi^b
=e^{2A(r_*)}\left[-f(r_*)dt^2+dx^idx^i\right]
 \label{5.4}\ee
 The effective speed of light on the 3-brane can be read from (\ref{5.4}) and is
\be
c^2_{eff}~=~f(r_*)~\leq ~c_{UV}^2
 \label{5.5}\ee
Near the bulk horizon, $f(r_h)=0$, the world-volume theory approaches the ``Carolean limit":  its effective speed of light vanishes $c_{eff}\to 0$.
In this limit,  different points in space cannot communicate anymore, and physics becomes 1-dimensional. Interestingly, the relevance of an analogous effect has been debated recently
in holographic applications to condensed matter physics, \cite{liu} and has been argued to be behind the critical behavior in holographic models for strange metals in the context of Schr\"odinger holography, \cite{kkp}. Unlike this, here the two-dimensional metric in the $(r,t)$ plane becames flat instead of AdS$_2$.

 In more complicated embeddings (like the brane moving with constant velocity in an internal dimension), the effective speed of light on the brane depends
on more data, both the bulk fields and the motion parameters,
\be
c^2_{eff}=f(r)-V^2
 \label{5.6}\ee
For a brane that stretches along the radial direction, the world-volume theory
has a world-volume horizon at $r_w$ where the effective speed of light vanishes, \cite{gkmmn,karch}
\be
c^2_{eff}=f(r_w)-V^2=0
 \label{5.7}\ee
In such a case LI is broken like in a black-hole background

It is instructive to review the breaking of LI as observed and interpreted by the inhabitants of the 3-brane world.
In the simplest first example, there is no signal of a gravitational field on the brane as it is flat. However the gravitional field is in the extra dimensions
and  it is responsible for the modified speed of light on the brane. This is indeed a modification of the gravitational force. If a scalar is also present,
it would give an extra contribution to the speed of light and therefore to LV.

In the second example of a brane that stretches in the radial direction, gravity is now visible on the brane, as the induced metric is that of the black-hole.
Again however the metric (and therefore the LV it induces) do not fit the matter densities that exist on the brane. Indeed the black hole metric is
due to the  motion of the brane in the extra dimensions. A similar idea led to mirage cosmology on branes in \cite{mirage}.

\subsection{Holography and Lorentz-violation}

In the setup of the previous section, if the metric (\ref{5.1}) becomes asymptotically AdS$_5$ near the boundary at
 $r=r_{\infty}$\footnote{ This means that $e^{A(r)}\sim {\ell\over r-r_{\infty}}$.},
we may have a holographically dual interpretation.

The bulk grvitational theory is dual to a boundary (3+1)-dimensional  large-N theory that we will denote by $\Theta_N$.
In particular the bulk metric (\ref{5.1}) corresponds to a concrete state of the theory $\Theta_N$.
The presence of the blackness factor in the metric (\ref{5.1}) signals that the state in question has a non-trivial energy density.
The presence of a horizon in (\ref{5.1}) signals that the state of $\Theta_N$ is thermal.

The theory $\Theta_N$ is coupled to the (3+1)-dimensional photon theory $\Theta_{ph}$ that lives on the 3-brane.
This theory has an {\cal O}(1) number of degrees of freedom and is weakly coupled. It interacts with the $\Theta_N$ theory via massive fields
that in our description have been integrated out to provide a coupling between the $\Theta_{ph}$ theory and  the stress tensor of the
 $\Theta_N$ theory (the bulk metric).
 The {\cal O}(1) number of degrees of freedom of the 3-brane indicates that it can be considered as a probe embedded in  the bulk gravitational
 background and its presence does not backreact on the bulk metric and other fields.

The position of the probe 3-brane in the radial direction corresponds to the RG scale of the $\Theta_{ph}$ theory.
If the brane is put at $r=r_{\infty}$, this would correspond to the bare action of the theory in the UV.
This is why the speed of light associated with $f(r_{\infty})$ was named $c_{UV}$.
Different positions in the radial direction correspond to different energy scales.
At the two derivative level the natutral connection between the radial coordinate  and the RG  energy scale $E$ is
\be
E(r)\equiv e^{A(r)}
 \label{5.8}\ee
The gravitional equations gurantee that $E$, defined as above, diverges at the UV boundary and monotonically decreases untill the IR end of space.
Moreover, near an asymptotically AdS$_5$ boundary, it agrees with the identification stemming from the AdS/CFT correspondence.

According to the above, the effective speed of light in the theory  $\Theta_{ph}$ given in (\ref{5.5}) should be viewed as the speed of light after one has included the quantum corrections due to the coupling to the $\Theta_N$ theory.
Because of (\ref{5.8}), the effective speed of light of $\Theta_{ph}$ is energy dependent,
\be
c_{eff}^2=f(r_*)=f(r_*(E))
 \label{5.9}\ee
Renormalization due to the $\Theta_N$ theory makes $c_{eff}$ decrease with energy.
In the most simple example, that of the AdS-Schwarschild geometry describing the thermal vacuum of the (3+1)-dimensional CFT we have
\be
e^A={\ell\over r}={E\over \ell}\sp f(r)=c_{UV}^2\left(1-(\pi Tr)^4\right)
 \label{5.10}\ee
which eventually gives
\be
c_{eff}=c_{UV}\sqrt{1-\left({\pi T\over E}\right)^4}
 \label{5.11}\ee
This is the behavior expected in a CFT where the coupling between $\Theta_N$ and $\Theta_{ph}$ is due to the stress tensor.

\FIGURE[t]{
\includegraphics[width=7cm]{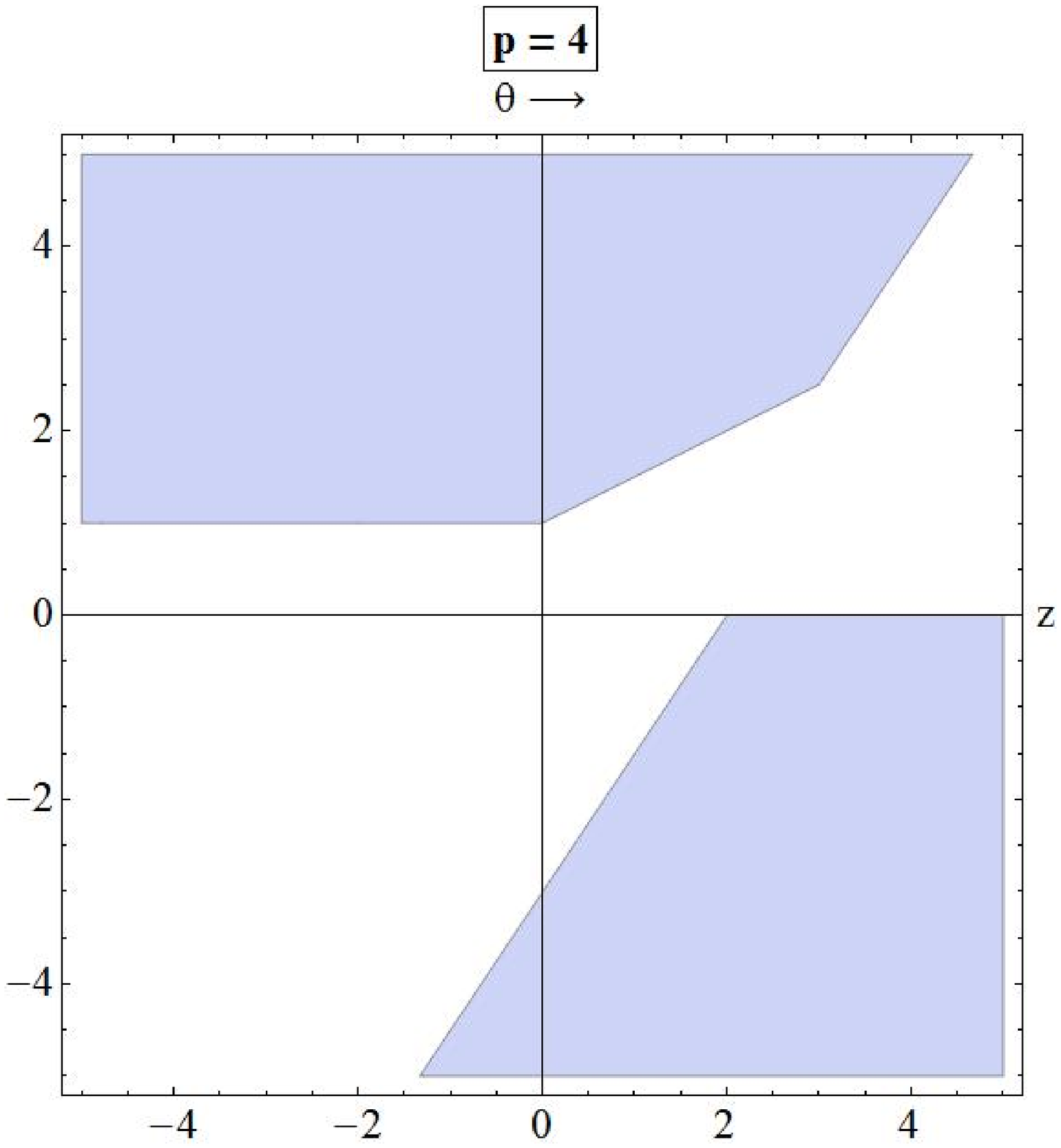}\includegraphics[width=7cm]{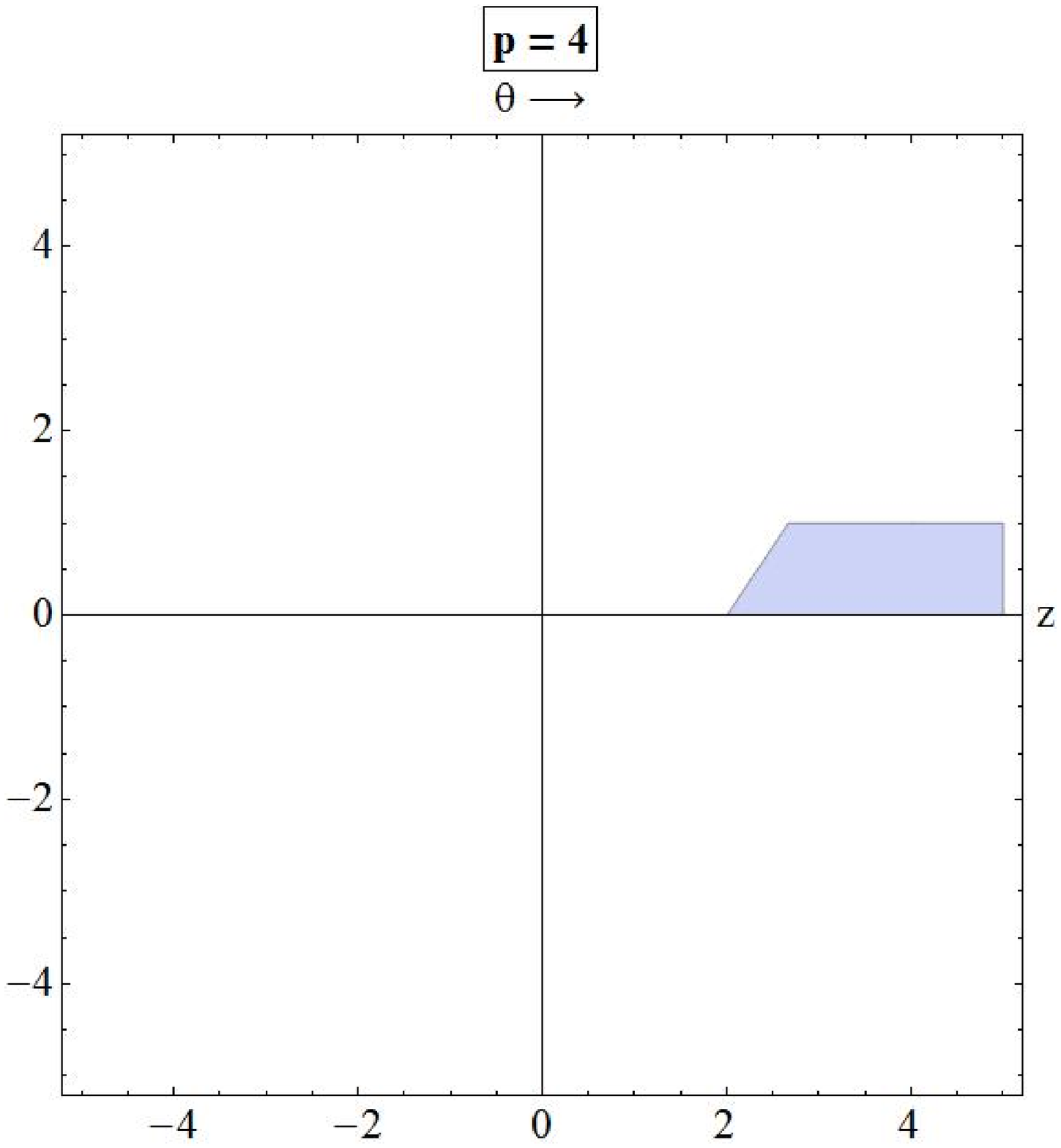}
\caption{Left:gappless generalized Lifshitz geometries in the $(z,\theta)$ plane. Right: gapped generalized Lifshitz geometries in the $(z,\theta)$ plane. All these geometries pass the Gubser bound tests in (\ref{a1.5}).}\label{fig6}}

Lorentz invariance is also broken by charge distributions.
It was argued in \cite{cgkkm} that although generically charge densities induce
a trivial (gapped) behavior at low energy/temperature, there are special cases where
there are non-trivial IR fixed points (quantum critical points) where the theory is
scale invariant albeit in a generic LV fashion.
The most general such metric is of the form, \cite{cgkkm,gk}
\be
ds^2=u^{\theta}\left[-{dt^2\over u^{2z}}+{b_0~du^2+dx^idx^i\over u^2}\right]
 \label{5.12}\ee
and is covariant under a generalized Lifshitz scaling symmetry
\be
 t\to \l^{z}t\sp u\to \l u\sp x^i\to \l x^i\sp ds^2\to \l^{-\theta} ds^2
 \label{5.13}\ee
with dynamical exponent $z$.
The exponent $\theta$ controls the transformation of the metric and is responsible for
 the non-standard scaling of physical quantities
like entropy.  It is known as the hyperscaling violation exponent.\footnote{It was shown in
\cite{cgkkm} that several of these geometries can be obtained from AdS or Lifshitz higher-dimensional
 geometries via standard dimensional reduction. This explains the modified
 scaling.}
These geometries have a scalar curvature
\be
R=-{3\theta^2-4(z+3)\theta+2(z^2+3z+6)\over b_0}~u^{-\theta}
\ee
The geometries are flat when $\theta=2$ and $z=0,1$. The $(\theta=0,z=1)$ geometry is in Ridler coordinates.
 The geometry is Ricci flat when $\theta=4$ and $z=3$.
The three previous special solutions  violate the Gubser bound conditions in (\ref{a1.5}).
The scalar curvature is constant when $\theta=0$ (pure Lifshitz case, \cite{kach}).
The figure \ref{fig6} displays the constraints of the Gubser bound  in (\ref{a1.5}) for gapless and gapped geometries.

We will do a radial redefinition
\be
u=(2-z)~r^{1\over 2-z}
 \label{5.14}\ee
together with a rescaling of $t,x^i$ to bring the metric to the form (\ref{5.1})
\be
ds^2\sim r^{\theta-2\over 2-z}\left[-f(r)dt^2+{dr^2\over f(r)}+dx^idx^i\right]\sp f(r)=f_0~r^{2{1-z\over 2-z}}
 \label{5.15}\ee
The case $z=2$ is special and must be dealt with separately.
In that case $u=e^r$ and the metric becomes
\be
ds^2\sim e^{(\theta-2)r}\left[-f(r)dt^2+{dr^2\over f(r)}+dx^idx^i\right]\sp f(r)\sim e^{-2r}
 \label{5.15a}\ee
The energy scale is given by the scaling of the $g_{tt}$ component of the metric. For a standard Lifshitz case (without violation of hyperscaling, $\theta=0$) we have
\be
E\simeq {u^{-z}}\simeq r^{-{z\over 2-z}}
 \label{5.16}\ee
However, when hyperscaling violations are present this relation becomes
\be
E\simeq u^{{\theta-2z}\over 2}\simeq {r^{\theta-2z\over 2(2-z)}}
 \label{5.17}\ee
We can therefore read the effective brane ``speed of light" and its dependence on energy as
\be
c_{eff}~~\sim~~ \sqrt{f}~~\sim~~ r^{{1-z\over 2-z}}~~\sim~~ { E^{2(z-1)\over 2z-\theta}}
 \label{5.18}
 \ee
This relation is also valid for $z=2$.
 In the case where the hyperscale violation is absent $\theta=0$, the scaling of the effective speed or light follows from Lifshitz symmetry.
 Indeed the relevant term in the action (\ref{4.7}) is
 \be
S_c=\int dt ~ d^{d}x\left[-c_{eff}^2~~\phi~\square\phi\right]
 \label{5.19}\ee
 Taking into account the scaling derived in (\ref{4.7}) the dimension of $c_{eff}$ is
 $[c_{eff}]=2(z-1)$ while the dimension of the energy $[E]=z$, which are compatible with the
 scaling relation in (\ref{5.18}) for $\theta=0$.

The constraints analyzed in \cite{cgkkm}, translated into $\theta,z$ imply that always
\be
  {z- 1\over 2z-\theta}>0\sp  c_{eff}\to \infty ~~~{\rm as}~~~E\to \infty
  \ee
 and vanishes at $E\to 0$.
This is consistent with a UV Lifshitz theory flowing towards the IR.
In the AdS case $z=1$, the effective speed of light is constant.
Under a general RG flow, (\ref{5.18}) provides the UV and IR asymptotics (with in general,
 different $z,\theta$) and some interpolating behavior in-between that encodes the RG flow.

We should mention that in the presence of regular horizons the theory $\Theta_N$ is
in a state with notrivial ${\cal O}(N^2)$ entropy.
The scaling geometries (\ref{5.12})  on the other hand, when $z\not= 1$ have zero
 entropy generalizing the example of \cite{gubser}.

To summarize,
\begin{itemize}

\item Braneworlds provide a natural setup for varying the speed
of light in QFT as well as providing other sources of LV.

\item  Most string theory constructions of the standard model use this paradigm,
 namely a SM  that is  composed out of  some D-branes in a 10d bulk, \cite{kir2,anastasopoulos}.
 For a supressed Lorentz violation in this type of brane configurations, either there should be very weak bulk fields
 or the branes should be close to each other and near the boundary. Put differently, different brane stacks (and therefore hidden sectors)
 are expected to have different speeds of light
in the presence of generic bulk fields.

\item  Generic bulk fields produced different speeds of light on
 different SM branes (that is  always smaller than the speed of gravitons in bulk)

\item  { The speed of light increases with energy,} $c\sim E^a$ but the exponent depends on the details of the theory.
We have calculated it via holography from the generic scaling solutions at finite density.

\item  {All of the above have an alternative interpetation of a SM interacting with a large-N theory in a LV state.}
\end{itemize}

\section{Lorentz violation and dissipation}

A general effect in the presence of LV couplings is the phenomenon of energy dissipation via Cerenkov radiation, \cite{cerenkov}.
In the simplest case, a charged particle moving faster than the speed of light in a given medium radiates photons.
This entails energy loss that is calculable at weak coupling.
It is a very important phenomenon that provides stringest constraints to LV couplings in many different contexts, especially
when applied to high energy cosmic rays \cite{Coleman,nm,ms}.
It is well undestood that in weakly coupled theories energy dissipation is unavoidable in LV contexts.

The fact that dissipation  generically follows LV can be seen in a simple class of examples with rotational symmetry intact. In that case, mass shell conditions can be parametrized by a single paramater, the speed of propagation. Assume three massless particles with distinct speeds, $c_a\sqrt{\vec P_a\vec P_a}=P^0_a$ and a three point vertex that couples them.
Assume that $c_3$ is the smalest of the speeds.
Energy and momentum conservation for the decay translates to $3\to 1+2$ is $P_3^{\mu}=P_1^{\mu}+P_2^{\mu}$ and we obtain
as a solution
\be
|\vec P^1|-{c_1c_2-c_3^2\cos\theta\pm\sqrt{(c_3^2\cos\theta-c_1c_2)^2-(c_1^2-c_3)^2(c_2^2-c_3)^2}\over c_1^2-c_3^2}|\vec P^2|=0
\label{a6}\ee
where $\vec P^1\cdot \vec P^2=|\vec P^1||\vec P^2|\cos\theta$.
(\ref{a6}) has a moduli space of solutions for a range of $\theta$, as $(c_1^2-c_3^2)(c_2^2-c_3^2)>0$ by assumption.
The moduli space is important because in the collinear case $cos\theta=1$ there is always a solution independent on the values of $c_i$,  but this is not enough for dissipation
when for example $c_1=c_2=c_3$.
For any other ordering of the speeds, the decay process must be adjusted appropriately.
Therefore, in the generic case, Cerenkov dissipation is always present.

The rate of energy loss depends crucially on the interaction.
Typically  it has the scaling form
\be
{dE\over dt}\sim E^a
\label{6.1}
\ee
with the power $a$ depending on the interaction.
For example, in the standard case and a medium with a reflaction constant independent of frequency $a=2$, \cite{cerenkov} while in the recent calculations for superluminal neutrinos,
 motivated by the initial OPERA results, the radiated energy is in the form of photons and $e^+e^-$ pairs with $a=3$, \cite{cohen}.
Different  types of gravitational Cerenkov radiation computed in the context of the Einstein-Aether theory, \cite{ms}, gives  $a=4,6,8$
depending on the type of process.

The question we would like to adress here is whether this phenomenon persists generically at strong coupling.
We will present a class of examples of energy dissipation from the holographic description of strongly coupled
QFTs that have been developped for applications to the physics of the quark gluon plasma, \cite{gubserrev}, and we will able to give the most general formula for energy loss
in scaling LV backrounds.

The particles in question are strongly interacting particles that we will call collectively ``quarks", motivated
by  the prototype situation. Such particles carry behind them a tube of flux that in the holographic description
is described by a fundamental string. There are several constraints for this description to be reliable, including the property that they are heavy.
The complete set of such constraints have been detailed in \cite{dress} and we will discuss them in detail later.
The motion of such a quark, or string in a bulk metric is described by the Nambu-Goto action,
 which is given by the product of the string tension times the area of its world-sheet,
\be
S_{\rm string}=-{1\over 2\pi\ell_s^2}\int d^2\sigma~\sqrt{\hat g}\sp \hat g_{ab}=G_{\m\n}{\pa X^{\m}\over \pa \sigma^a}
{\pa X^{\n}\over \pa \sigma^b}
\label{6.2}\ee
with $X^{\m}$ the embedding coordinates of the string in the bulk, $\sigma^a$ the worl-sheet coordinates and $G_{\m\n}$ the bulk metric in the string frame.

We will consider a quark that is moving in a bulk metric (in the string frame)  of the form
\be
ds^2=b(r)^2\left[{dr^2\over f(r)}-f(r)dt^2+d\vec x\cdot d\vec x\right]
\label{6.3}\ee
We will force the quark by an external source to move with a constant velocity $v$, and we will find its shape by
solving the Nambu-Goto equations of motion.
We will make the ansatz
\be
x^1=vt+\xi(r)\sp x^{2,3}=0 \sp \sigma^1=t\sp \sigma^2=r
\label{6.4}\ee
where we have also fixed the static gauge by choosing the values of $\sigma^{1,2}$.
The profile of the trailing string is given by the function $\xi(r)$ that is shown in figure \ref{fig1}.

\FIGURE{
\includegraphics[width=12cm]{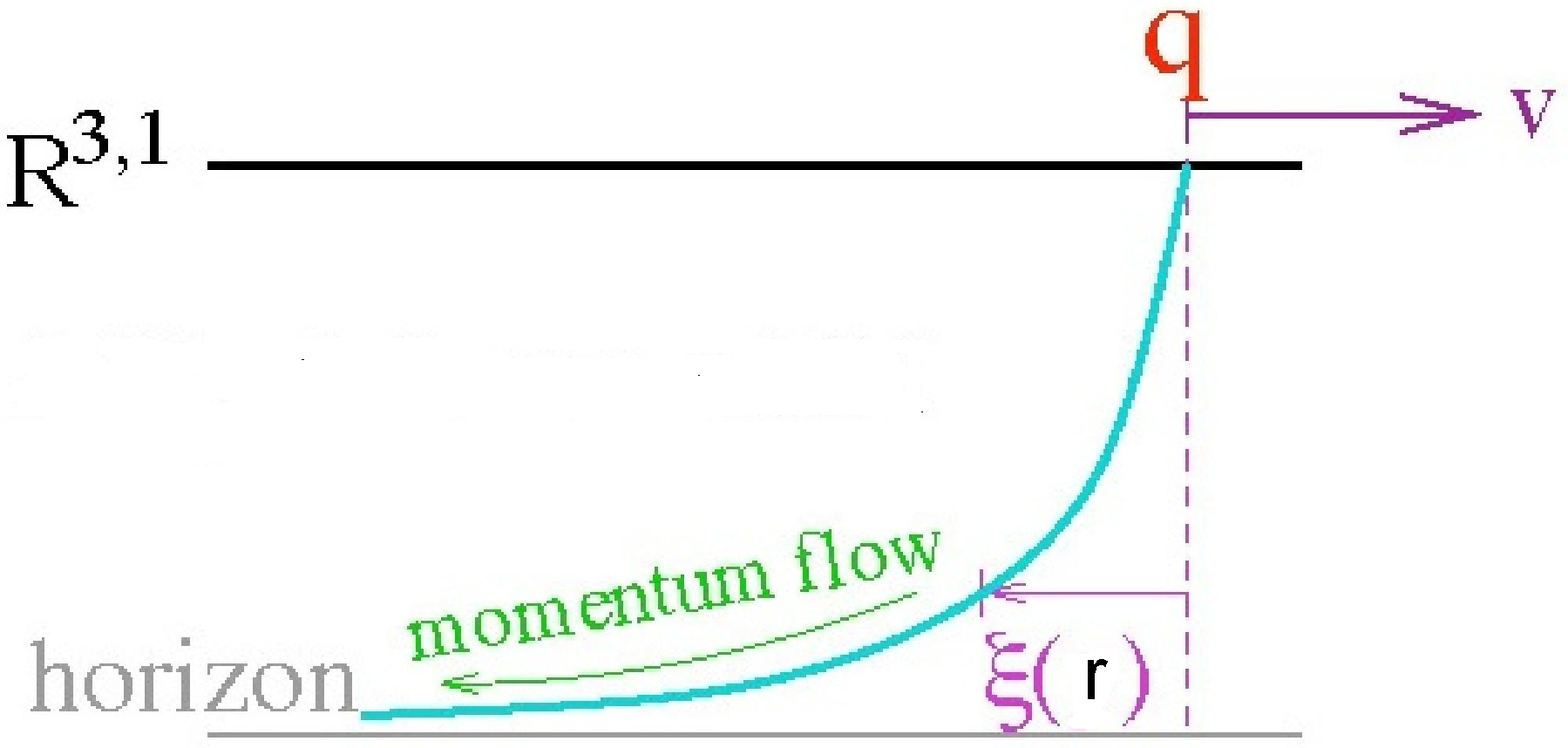}\hspace{1cm}
\caption{The trailing string profile.}\label{fig1}}
The induced metric can be calculated to be
\be
g_{\alpha\beta} = b^2(r) \left(\begin{array}{cc}v^2 - f(r)& v \xi'(r)\\ v \xi'(r) & f(r)^{-1} + \xi^{'2}\end{array}\right),
\label{6.5}\ee
and the corresponding Nambu-Goto action:
\begin{equation}
S = -\frac{1}{2\pi \ls^2} \int dt dr ~ b^2(r)\sqrt{1 - {v^2\over f(r)}+ f(r) \xi^{'2}(r)} \;.
\label{6.6}\end{equation}
Since $S$ does not depend on $\xi$ but only its derivative, the
conjugate momentum $\pi_\xi$ is conserved,
\be
\pi_\xi = -{1\over 2\pi \ell_s^2}{ b^2(r) f(r) \xi'(r) \over \sqrt{1 - {v^2\over f(r)} + f(r) \xi^{'2}(r)}}
\label{6.7}\ee
from which we obtain
\be
\xi'(r)={C\over f(r)}\sqrt{f(r)-v^2\over b^4(r)f(r)-C^2}\sp C=2\pi\ls^2 \pi_\xi
\label{6.8}\ee
The constant C is fixed by the following argument: the numerator inside the square root of (\ref{6.7}) changes sign and becomes negative at $r=r_s$ where
\be
f(r_s)=v^2
\label{6.9}\ee
The denominator will change sign at some other radial position for arbitrary $C$. For the solution to be well defined, it must change sign at the same $r=r_s$.
This fixes $C$ and $\pi_\xi$ to be
\be
\pi_\xi = - {b^2(r_s) \sqrt{f(r_s)}\over 2\pi \ell_s^2}=- {b^2(r_s)\over 2\pi \ell_s^2}v\sp  C\equiv v \, b^2(r_s).
\label{6.10}\ee

The {\em drag force} on the quark, due to the motion inside the metric  can be
determined by calculating the momentum that is lost
by flowing from the string to the horizon, which results in:
\be
 F_{\rm drag} = \pi_\xi = - \frac{v\, b^2(r_s)}{2\pi \ls^2},
\label{6.11}\ee
where we have replaced $f(r_s)$ by $v^2$ in the last equality.

One defines the momentum {\em friction coefficient} $\eta$ as the characteristic attenuation constant for the momentum of a (heavy) particle :
\be
F_{\rm drag}=-{\eta}v,
\label{6.12}\ee
With this definition we obtain:
\be\label{6.13}
\eta =  { b^2(r_s) \over 2\pi \ell_s^2 }.
\ee
The coordinate value $r=r_s$ is
 a horizon for the induced world-sheet
metric, \cite{teaney,gkmmn}. To show this we change coordinates to diagonalize the
induced metric, by means of the reparametrization:
\be
t=\tau+\zeta(r), \qquad \zeta'={v\xi'\over
f-v^2}={Cv\over f(r)\sqrt{(f(r)-v^2)( b^4f-C^2)}}. \label{6.14} \ee
In these coordinates the induced metric is
\be
ds^2=b^2\left[-(f(r)-v^2)d\tau^2+{b^4\over
(b^4f-C^2)}dr^2\right].
\label{6.15}\end{equation}
The coefficient of $d\tau^2$ vanishes at $r_s$, so this point corresponds to a world-sheet black-hole horizon.
Since $f(r)$ runs between $0$ and $c^2$ as $0<r<r_h$, by definition (\ref{6.9})
the world-sheet horizon is
always outside the bulk black hole horizon, and it coincides with it
only in the limit $v\to 0$. In the opposite limit, $v\to c$, $r_s$
asymptotes towards the boundary $r=0$.

The Hawking temperature associated to the black hole metric (\ref{6.15}) is found as usual,
 by expanding around $r=r_s$ and demanding regularity of the Euclidean geometry. The resulting temperature is:
\be \label{6.16}
 T_{s} \equiv {1\over 4\pi}\sqrt{f(r_s)f'(r_s)\left[{4b'(r_s)\over b(r_s)}+{f'(r_s)\over f(r_s)}\right]}.
\ee

In the conformal limit, where the background solution reduces to $AdS$-Schwarzschild,
\be
b(r)={\ell\over r}\sp f=c^2\left[1-(\pi Tr)^4\right]
\label{6.17}\ee
the world-sheet temperature and horizon position are
simply given by:
\be\label{6.18}
T_s^{conf} = {T\over \sqrt{\gamma}}, \qquad r_s^{conf} = {1\over \pi \sqrt{\gamma}\, T}.
\ee

 In the conformal case, $\eta$ can be defined in a relativistic fashion and is independent of $p$,
\be\label{6.19}
\eta^{conf}_D={ \pi \sqrt{\l}~T^2 \over 2M_q}\sp {dp\over dt}=-\eta_D^{conf} p
\ee
where $\lambda=(\ell/\ell_s)^4$ is the fixed 't Hooft coupling of ${\cal N}=4$ sYM.
This is not anymore so in the general case, where $\eta $
is velocity/momentum dependent.

We would like now to interpret the calculation in the light of our previous discussion.
First we observe that the energy loss is possible only if the bulk metric violates LI and the blackness function $f$ is variable.
Indeed, if $f=constant$, the string is hanging straight down and there is no energy loss.\footnote{There is an exception in confining vacuua. There, although $f$ is constant,
there is a minimum of the scale factor in the string frame. This alows a non-trvial solution (that is related to the non-trivial Wilson line configuration of the confining saddle-point), \cite{nit}. In a sense this is the 1/2 of the confining Wilson line saddle-point solution.}

The next observation is that  all portions of the trailing string move with  the same velocity, namely $v$.
The local ``velocity of light" at radial position $r$, is given by $c_{eff}^2=\sqrt{f(r)}$ and varies from the standard speed of light $c$ down to zero (at the bulk horizon).
The portion of the string below the world sheet horizon at $r=r_s$ is locally superluminal.
It therefore dissipates energy (that in this example is provided by the source).
This is therefore a strong coupling analogue of Cerenkov radiation. There are however differences from the weak coupling analogue. Here the energy is not radiated in the bulk
\footnote{There will also be conventional gravitational radiation in the bulk, it is however supressed at large $N$ by a $1/N^2$ factor.
There is also another form of Cerenkov radiation that the quark emits that is in the form of mesons, \cite{mateos}. This is again supressed at large N.}
but rather is dissipated through the string.

The onshell action of the string can be evaluated to be
\be
S=
\frac{1}{2\pi \ls^2} \int dt\int_{r_b}^{r_h} dr ~ b^4(r)\sqrt{f-v^2\over b^4f-C^2} \;.
\label{a1.2}\ee
where $r_b$ is a shifted boundary coordinate (the position of the string end-point), while $r_h$ is the position of the bulk horizon.
From this we can read the Lagrangian for the strongly interacting particle as
\be
L=\frac{1}{2\pi \ls^2} \int_{r_b}^{r_h} dr ~ b^4(r)\sqrt{f-v^2\over b^4f-C^2}
\label{a1.3}\ee
In AdS$_5$ case with $b(r)={\ell\over r}$, $f(r)=1-(\pi Tr)^4$, we obtain
\be
L_{AdS}=\frac{\ell^2\sqrt{1-v^2}}{2\pi \ls^2} \int_{r_b}^{\infty} {dr\over r^2}=M\sqrt{1-v^2}\sp M\equiv {\ell^2\over 2\pi\ell_s^2 r_b}={\sqrt{\lambda}\over 2\pi r_b}
\label{a1.4}\ee
This is indeed the relativistic energy of a point particle with velocity $v$.

We will now proceed to compute the energy loss in the generalized scaling metrics at finite density\footnote{The fact that zero temperature Lifshitz metrics entail energy loss was independently observed by D. Tong. The drag force in Lifhitz space-times was also calculated in \cite{fadafan}}
introduced in the previous section that we reproduce here
\be
ds^2\sim \left({r\over \ell}\right)^{\theta-2\over 2-z}\left[-f(r)dt^2+{dr^2\over f(r)}+dx^idx^i\right]\sp f(r)= f_0 \left({r\over \ell}\right)^{2{1-z\over 2-z}}
 \label{6.20}\ee
 These metrics are describing either UV or IR asymptotics with a non-relaticisitic scaling symmetry and hyperscaling violation.
 The boundary is at $r=0$ when ${\theta-2\over z-2}>0$ and at $r=\infty$ when ${\theta-2\over z-2}<0$.

The turning point is determined by
\be
f_0~\left({r_s\over \ell}\right)^{2(1-z)\over 2-z}=v^2~~~~\to~~~~ {r_s\over \ell}=\left({v\over \sqrt{f_0}}\right)^{2-z\over 1-z}
 \label{6.21}\ee
 Unlike the AdS case, there is no limit velocity here as $f$ diverges at the boundary. The same remains true if you consider Lifshitz geometries at finite temeperature (Lifshitz black holes).

\be
F_{\rm drag}=-{v b^2(r_s)\over 2\pi\ls^2}=-{v\over 2\pi\ls^2} \left({r_s\over \ell}\right)^{\theta-2\over 2-z}=
-{\sqrt{f_0}\over 2\pi\ls^2} \left({v\over \sqrt{f_0}}\right)^{ z+1-\theta\over z-1}=-{v\over 2\pi\ls^2} \left({v\over \sqrt{f_0}}\right)^{ 2-\theta\over z-1}
 \label{6.22}\ee
 The equation above can be written in vectorial form as
\be
\vec F_{\rm drag}=-{\vec v\over 2\pi\ls^2} \left({|\vec v|\over \sqrt{f_0}}\right)^{ 2-\theta\over z-1}
 \label{6.22a}\ee

The Hawking temperature is given by
\be
4\pi T_s={2f_0\over |z-2|\ell}\sqrt{(z-1)(z+1-\theta)}\left({r_s\over \ell}\right)^{z\over z-2}=
{2f_0\over \ell}\left({r_b\over \ell}\right)^{z\over z-2}{\sqrt{(z-1)(z+1-\theta)}\over |z-2|}
{v}^{z\over z-1} \label{6.23}\ee

When $z=2$ the relation becomes
\be
T_s=2\sqrt{3-\theta}~{v^2\over 2\pi\ell}
 \label{6.24}\ee
 The energy function has been computed in appendix \ref{a} in (\ref{a1.10}), (\ref{a1.11}).
 The equation of motion of the particle is given by
 \be
 {d\over dt}{dL_{z,\theta}(v)\over dv}=F_{\rm drag}~~~\to~~~ M_{||}(v) {dv\over dt}=F_{\rm drag}
\ee
with $M_{||}$ defined in (\ref{7.36}).
In the low velocity regime, $v\to 0$, using the results of appendix \ref{a} we obtain

\be
\ell{dv\over dt}\simeq \Big|1+{2z-\theta\over z-2}\Big|f_0\left({r_b\over \ell}\right)^{\theta -1\over z-2}\left({v\over \sqrt{f_0}}\right)^{z+1-\theta\over z-1}
\ee
For a pure Lifshitz backround ($\theta=0$), it implies a velocity fall-off
\be
v(t)\sim t^{-{z-1\over 2}}
\ee
Note also that in the case $\theta={z+3\over 2}$, the energy loss rate is exponential.

In the large velocity regime $v\to \infty$, using the results of appendix \ref{a} we obtain instead
\be
\ell{dv\over dt}\sim~
\left({v\over \sqrt{f_0}}\right)^{3z+1-2\theta\over z-1}
\ee
Identifying the energy as $E=L_{z,\theta}(v)-v{dL_{z,\theta}\over dv}$ we obtain equivalently
\be
{dE\over dt}\sim E^{2z-\theta\over \theta-2}
\ee
This implies that the dissipation rate vanishes at high energies and the energy behaves as $E\sim (t_0-t)^{z+3\over 2}$ in the Lifshitz case $(\theta=0)$.
However, as we will show in the next section the sign is changed and instead of dissipation it is an energy gain.
This is unusual and signals an instability.

The previous calculations show that there is a general range of dissipation rates that run from very slow (when ${z+1-\theta\over z-1}\ll 1$ or ${z+1-\theta\over z-1}\gg 1$
to very fast (exponential) when ${z+1-\theta\over z-1}\simeq 1$.
In the first case, such LV is well hidden from the dissipation observational window, that typically is a very sensitive test,
 especially for probes that arrive from very fast.

\subsection{On the validity of approximations}

There are two important approximations that are assumed in the description  of quark energy loss in terms of semicalssical string motion.
The first is that the world-sheet horizon must remain below the endpoint of the string (that has been positioned at $r=r_b$).
This implies ${r_s\over r_b}\simeq v^{z-2\over z-1}\ll 1$ if the boundary is at $r=\infty$, or   ${r_s\over r_b}\simeq v^{z-2\over z-1}\gg 1$ if the boundary is at $r=0$.
 We call this the ``horizon criterion".

Another essential approximation in our calculation is that we neglected
 the dependence of the particle action on acceleration. For this we must have that the acceleration is smaller than velocity.
A relevant ratio is that of the acceleration to velocity. As $v\to 0$ we calculate
\be
{\ell^2 \ddot v\over \ell \dot v}\sim v^{{(2-\theta)\over z-1}}
\ee
while as $v\to \infty$ we obtain instead
\be
{\ell^2 \ddot v\over \ell \dot v}\sim v^{{2(z+1-\theta)\over z-1}}
\ee
with ${(z+1-\theta)\over z-1}>0$ always.

We may now analyze the validity of these two approximations in the three regimes of appendix \ref{a}.

\begin{itemize}

\item{1a)} ${\theta-2\over z-2}>0$ and ${z-1\over z-2}>0$. Here the horizon criterion is satisfied at sufficiently high velocity.
On the other hand the acceleration criterion fails both at low and high velocities.
Therefore, this is a class of theories for which our energy loss calculation is unreliable in both asymptotic regimes.

\item{1b)}  ${\theta-2\over z-2}>0$ and ${z-1\over z-2}<0$. Here the horizon criterion is satisfied at sufficiently low velocity and the same applies to the acceleartion criterion.
Therefore the calculation in the low velocity regime is reliable here.

\item{2a)}  ${\theta-2\over z-2}<0$ and ${z-1\over z-2}>0$. Here the horizon criterion is satisfied at sufficiently low velocity and the same applies to the acceleartion criterion.
Therefore the calculation in the low velocity regime is also reliable here.

\end{itemize}

\section{Fluctuations and dissipation}

We consider a heavy particle which, in a first approximation, experiences a
uniform motion across a LV strongly coupled medium, with constant velocity $v$. Due
to the interactions with the strongly-coupled medium, the actual
trajectory of the particle  is expected to resemble Brownian motion. To
lowest order, the action for the external particle coupled to the
medium  can be assumed, classically, to be of the form:
\be\label{bound action} S[X(t)] = S_0 + \int d\tau X_\mu(\tau)
{\cal F}^\mu(\tau) \ee
where $S_0$ captures the action of the free particle and the last term involves the force to the particle from the backround.

\subsection{The Langevin evolution}

There is a standard way of deriving a Langevin evolution for such a particle, provided it is sufficiently heavy, and this is detailed in \cite{gkmmn}.
Here we quote the result for the Markovian limit in which the Langevin evolution is local in time with white noise,
\be\label{langeq2} {\delta S_0 \over \delta X_i(t)} + \eta^{ij}
\dot{X}_j(t) = \xi^i(t), \qquad \langle\xi^i(t)\xi^j(t')\rangle =
\kappa^{ij} \delta(t-t'),
\ee
with the self-diffusion and
friction coefficients given by:
\be \label{coeff} \kappa^{ij} =
\lim_{\omega \to 0} G_{sym}^{ij}(\omega); \qquad \; \eta^{ij}
\equiv \int_0^\infty d\tau\, \gamma^{ij}(\tau) = -\lim_{\omega
\to 0} { {\rm Im}\, G_R^{ij} (\omega) \over \omega}. \ee
where the correlators of the force have been defined and analyzed in appendix \ref{la}.

In the case of a system at equilibrium with a canonical ensemble
at temperature $T$, one has the following relation between the
Green's functions:
\be\label{thermal} G_{sym}(\omega) = -
\coth{\omega\over 2T}\, {\rm Im} \,G_R(\omega), \ee
 which using equation
(\ref{coeff}) leads to the Einstein relation $\kappa^{ij}= 2 T
\eta^{ij}$.
For such a thermal ensemble, the real-time correlators decay exponentially with a scale
set by the inverse temperature, therefore
the typical correlation time is $\tau_c \sim 1/T$.

 Determining and studying the Langevin correlators (\ref{correlators1})  and
the diffusion constants (\ref{langeq2}) will be the main purpose
of the rest of this paper.

We write explicitly the classical part, $\delta S_0
/\delta X(t)$ of the Langevin equation, in order to arrive at an
equation describing momentum diffusion. We start with the kinetic
action for the particle, derived in appendix \ref{a},
 \be
 S_0[X_\mu(\tau)] =- \int dt ~L(\dot X)
\label{7.28}\ee
We choose the gauge $\tau = X^0$, and obtain,
\be
S_0(\dot X^i=v^i)=-\int dt L_{z,\theta}(\vec v^2)
\label{7.29}\ee
with $L_{z,\theta}$ given in (\ref{a1.10}) and
 \be
 {\delta S_0\over \delta X^i(\tau)} = {d p_i\over dt} \sp {\rm with}\sp p_i \equiv -{\delta L_{z,\theta}(\vec v^2)\over \delta v^i}=-2L'(\vec v^2)v^i.
\label{7.30}\ee
where the prime stands for a derivative with respect to $\vec v^2$.

 Equation (\ref{langeq2}) becomes
the Langevin equation for momentum diffusion:
\be\label{langeq3}
{d p^i \over d t} = - \eta^{ij}(\vec{v}^{~2}) v^j + \xi^i(t),
\ee
The drag force calculated in the previous section, (\ref{6.22a}), is identified with the friction term on the right hand side
\be
\eta^{ij}={1\over 2\pi\ell_s^2}\left({\vec v^2\over f_0}\right)^{2-\theta\over 2(z-1)}~\delta^{ij}\equiv \eta(\vec v^2)~\delta^{ij}
\label{7.31}\ee

\paragraph{Linearized Langevin equations.}
For a generic particle trajectory, the Langevin equation
(\ref{langeq3}) is non-linear,
 due to the $v$-dependence in $\eta^{ij}$.
To put it in a form which allows for the holographic treatment in
terms of the trailing string fluctuations, it is convenient to
derive from equation (\ref{langeq3}) a linearized Langevin equation
for the fluctuations in the position around a trajectory with
uniform velocity, $\vec{X}(t) = \vec{v}t + \delta \vec{X}$. We therefore
 separate the longitudinal and transverse components
of the velocity fluctuations:
\be \dot{\vec{X}}(t) = \left(v +
\delta \dot{X}^\parl(t)\right){\vec{v} \over v} + \delta
\dot{\vec{X}}^\perp=\vec v+\delta \vec v\sp \vec v\cdot \delta\dot{\vec{X}}^\perp=0.
\label{7.32}\ee

The corresponding linearized expression
of the momentum reads:
\be \vec{p} =-2L'\vec v-2L' {\delta \dot{\vec X}^\perp}-(2L'+4\vec v^2 L''){\vec v\over v}  {\delta \dot{\vec X}^\parl}= \vec{p_0} + \delta
\vec{p},
\label{7.33}\ee
 The zeroth-order term is
$\vec{p}_0 =-2L'\vec v$, and the longitudinal and
transverse momentum fluctuations are given by:
\be \delta p^\parl =-(2L'+4\vec v^2 L'')
\delta \dot{X}^\parl, \qquad \delta p^\perp_i =-2L' \delta
\dot{X}^\perp_i.
\label{7.34}\ee
It is convenient to separate the longitudinal
and transverse components of the propagators, since,  as it will become
clear later on, the off-diagonal components vanish:
\be
G^{ij}(t) = G^\parl(t) {v^i v^j \over v^2} + G^\perp(t)
\left(\delta^{ij} - {v^i v^j \over v^2}\right)
\label{7.35}\ee
and the
corresponding decompositions for $\eta^{ij}$ and $\kappa^{ij}$ from (\ref{coeff}).
In particular from (\ref{7.6}), $\eta^{\perp}=\eta^{||}=\eta$.

Inserting these expressions in equation (\ref{langeq3}), we find to
zeroth order: \be\label{zeroth} {d p_0 \over d t} = -\eta^\parl
v , \ee and to first order in
$\delta \vec{X}$ the relativistic Langevin equations for
position fluctuations:
\bea
&& -(2L'+4v^2L'')\ddot{X}^\parallel = - \hat \eta^\parallel(v) \delta \dot{X}^\parallel + \xi^\parallel,
 \qquad \langle\xi^\parl(t) \xi^\parl(t') \rangle= \kappa^\parl \delta(t-t'), \label{langeq4-a}\\
&& -2L'\delta \ddot{X}^\perp =- \hat \eta^\perp(v) \delta
\dot{X}^\perp + \xi^\perp, \qquad \langle\xi^\perp(t)
\xi^\perp(t')\rangle = \kappa^\perp \delta(t-t')\label{langeq4-b}
\eea
where the friction coefficients $\hat \eta^{\parl,\perp}$ are
related to the coefficient $\eta$ by
\be\label{fri}
\hat \eta^\perp =\eta, \qquad \hat \eta^\parl =\eta+2v^2\eta'={z+1-\theta\over z-1}~\eta. \ee
Note that the equations involve velocity-dependent effective masses as
\be
M_{||}\equiv -(2L'+4v^2L'')=-{d^2L\over dv^2}\sp M_{\perp}\equiv -2L'=-{1\over v}{dL\over dv}
\label{7.36}\ee
The Langevin equations can be simplified as
\bea
&& \ddot{X}^\parallel = \tilde \eta^\parallel(v) \delta \dot{X}^\parallel + \tilde \xi^\parallel,
 \qquad \langle\tilde\xi^\parl(t) \tilde\xi^\parl(t') \rangle= \tilde\kappa^\parl \delta(t-t'), \label{langeq4-aa}\\
&& \delta \ddot{X}^\perp =\tilde \eta^\perp(v) \delta
\dot{X}^\perp + \tilde\xi^\perp, \qquad \langle\tilde\xi^\perp(t)
\tilde\xi^\perp(t')\rangle = \tilde\kappa^\perp \delta(t-t')\label{langeq4-bb}
\eea

\be\label{7.37}
\tilde \eta^\perp ={\eta\over 2L'}, \qquad \tilde \eta^\parl ={z+1-\theta\over z-1}~{\eta\over (2L'+4v^2L'')}. \ee

For comparison, the function that we obtain from the AdS-Schwarschild black-hole is $L_{z=1}={M}\sqrt{1-v^2}$, with M cutoff dependent (see (\ref{a1.4})). The existence of a maximal speed here is associated to the fact that the
blackness factor is finite near the AdS boundary.
This gives, \cite{gkmmn},
\be
M_{\perp}={M\over \sqrt{1-v^2}}\sp M_{||}={M\over (\sqrt{1-v^2})^3}
\ee

\FIGURE{
\includegraphics[width=7cm]{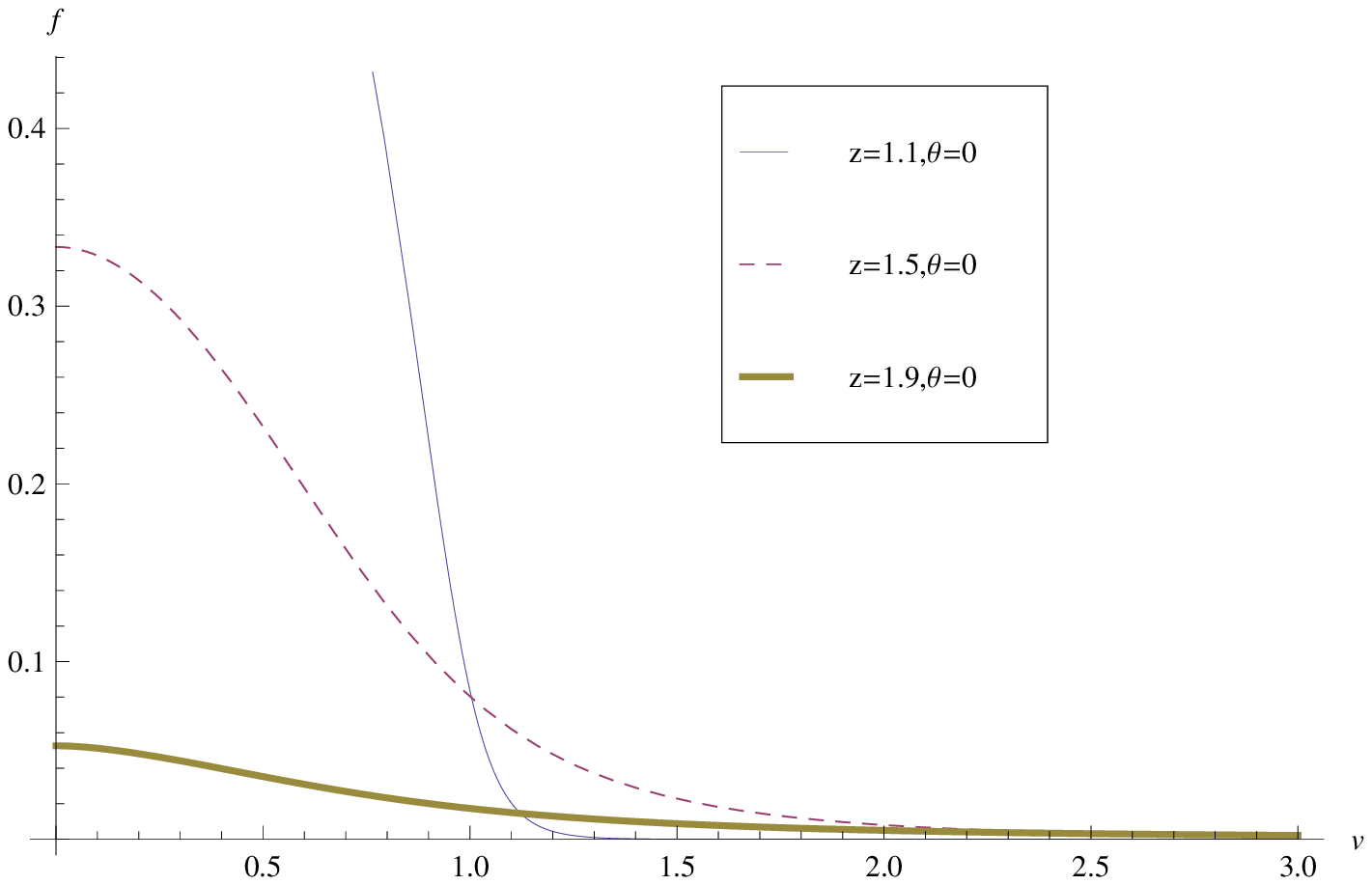}\includegraphics[width=7cm]{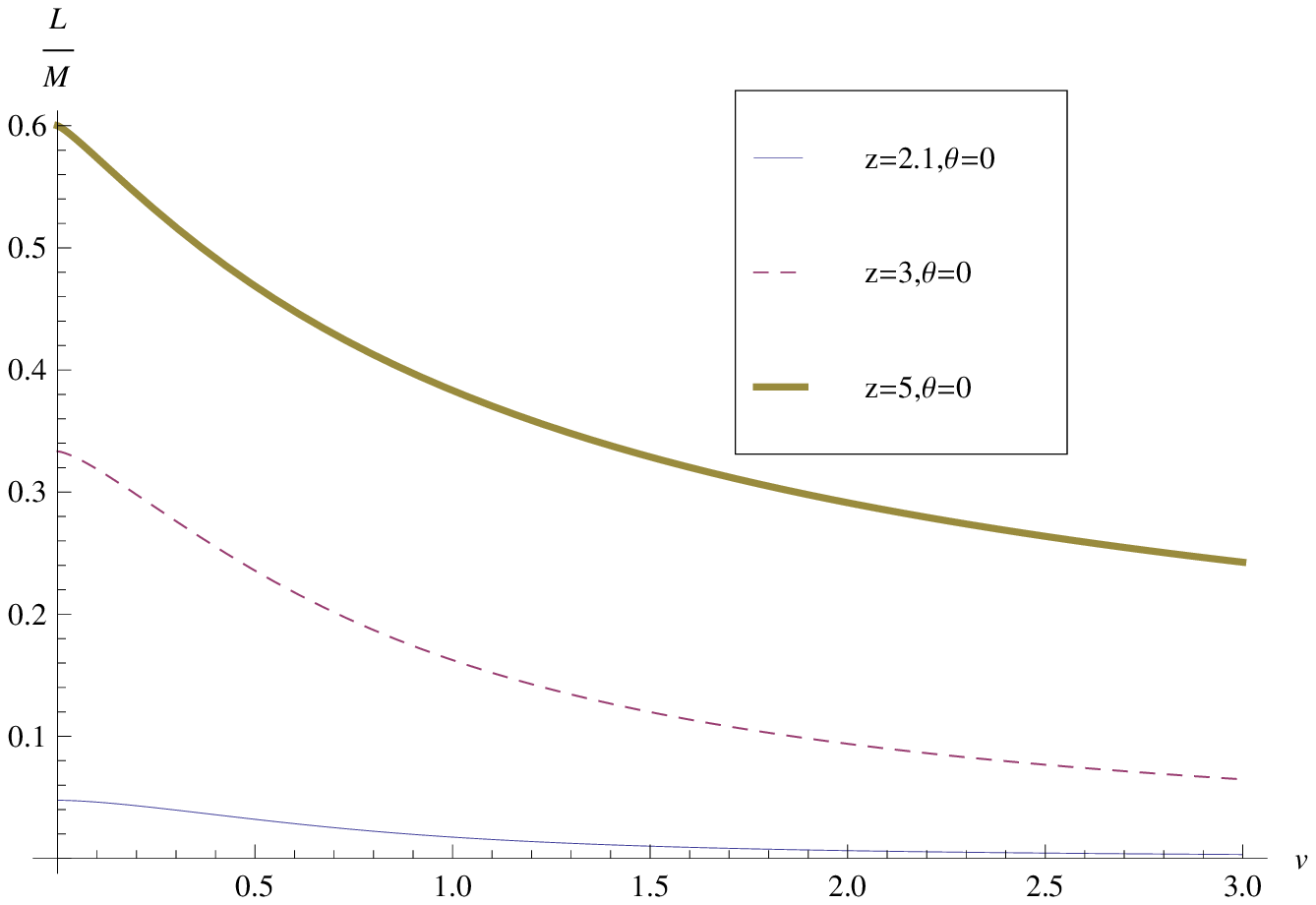}
\caption{Left: A plot of the function ${L\over M}=v^{z-\theta\over z-2}F[z,\theta;v^{-{\theta-2\over z-2}}]$ as  function of $1<z<2$ at $\theta=0$. Right: the same function for $z>2$ at $\theta=0$.}\label{fig2}}

\FIGURE[t]{
\includegraphics[width=7cm]{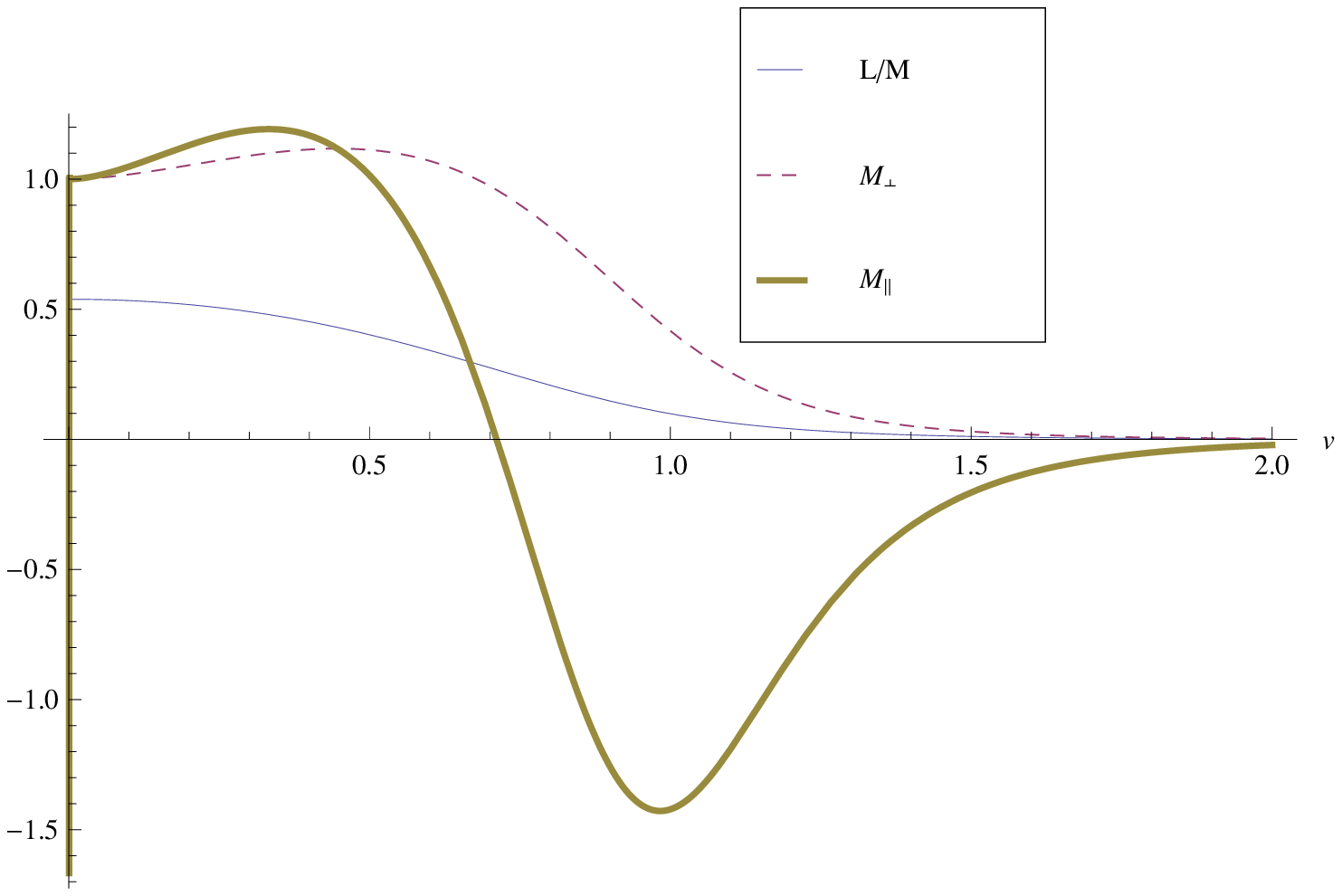}\includegraphics[width=7cm]{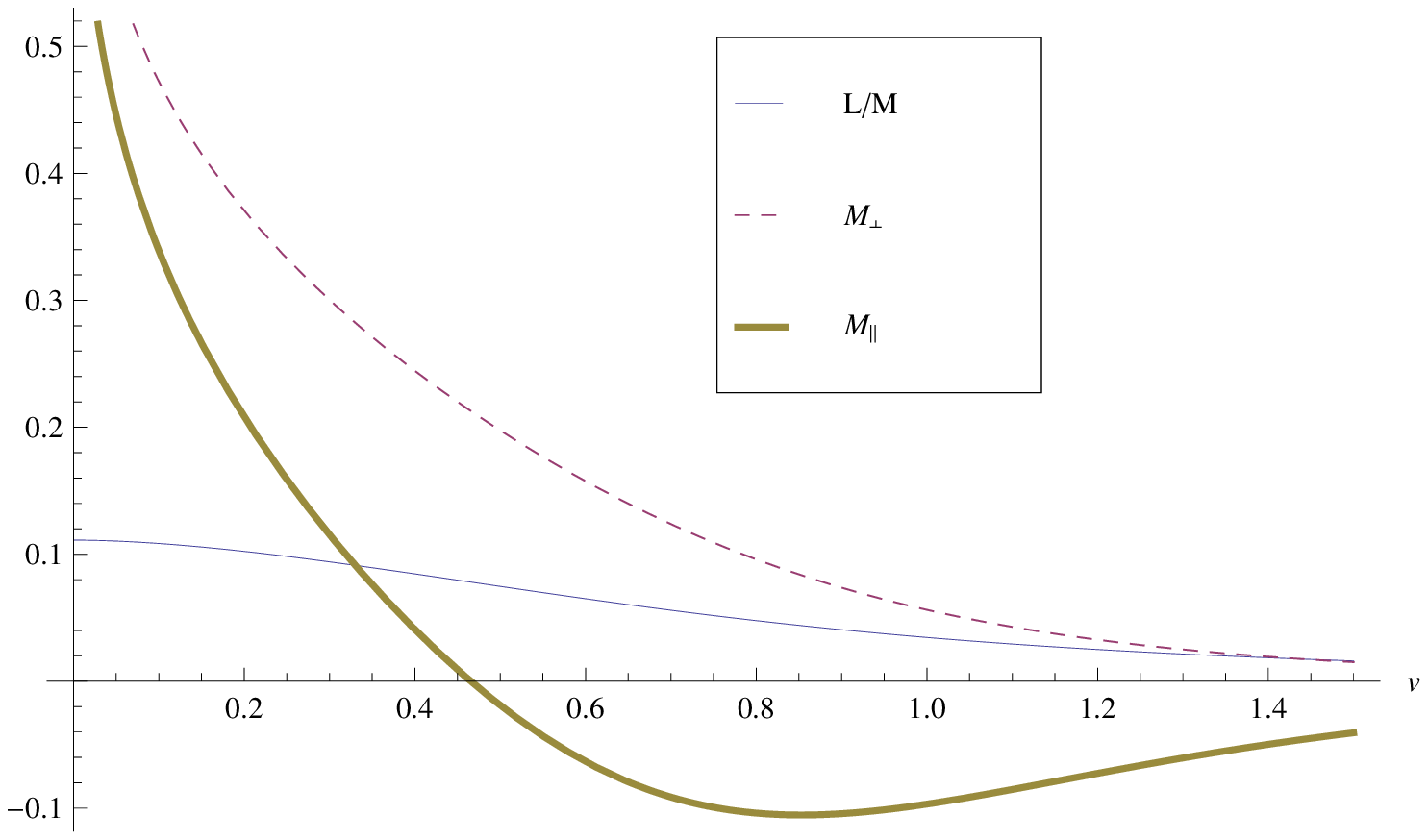}\hspace{1cm}
\caption{Left:A simultaneous plot of ${L\over M}$, $M_{||}$ and $M_{\perp}$ at $z=1.3$. Right: the same plot for $z=1.8$.}\label{fig3}}

\FIGURE[b]{
\includegraphics[width=7cm]{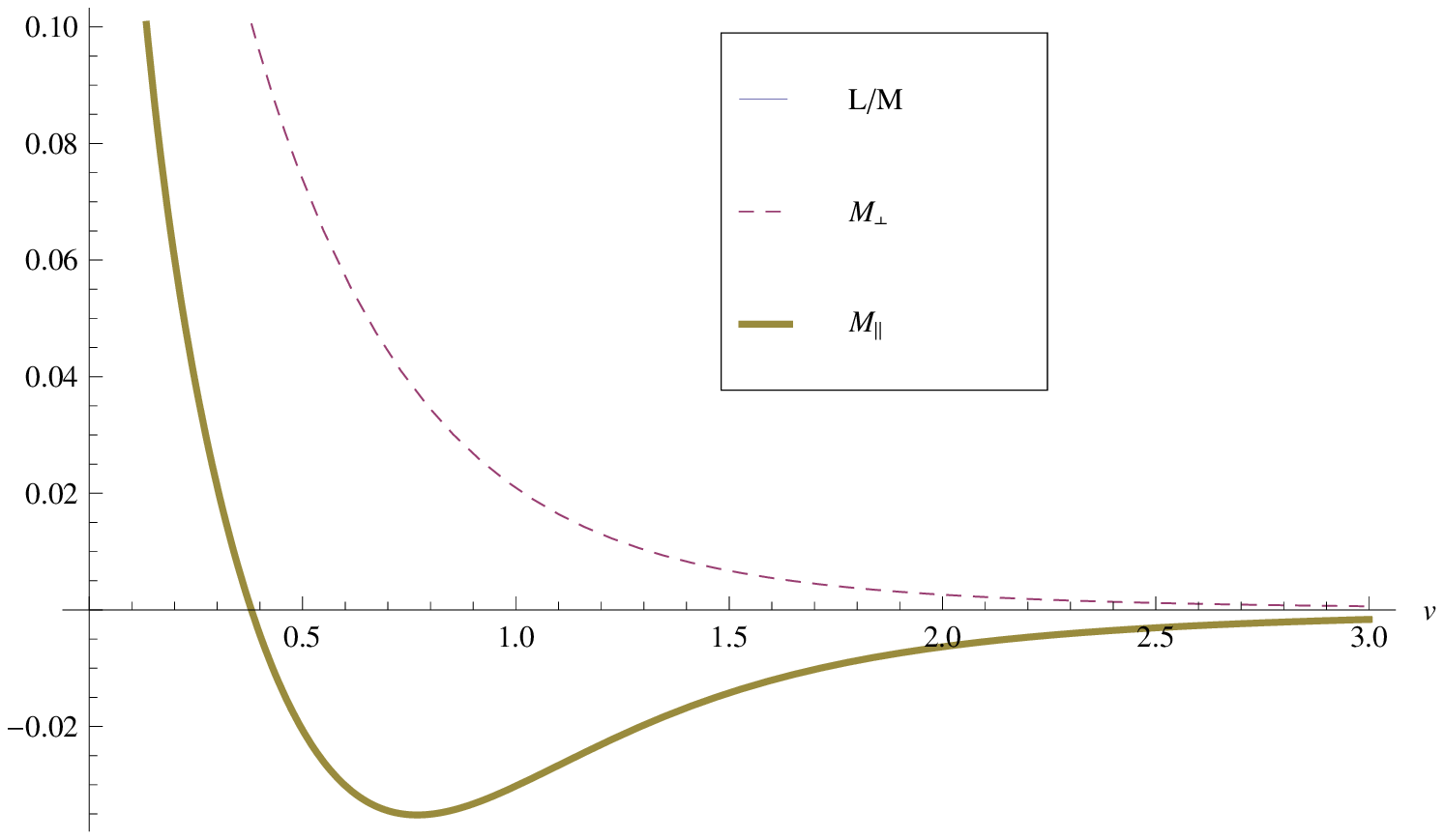}\includegraphics[width=7cm]{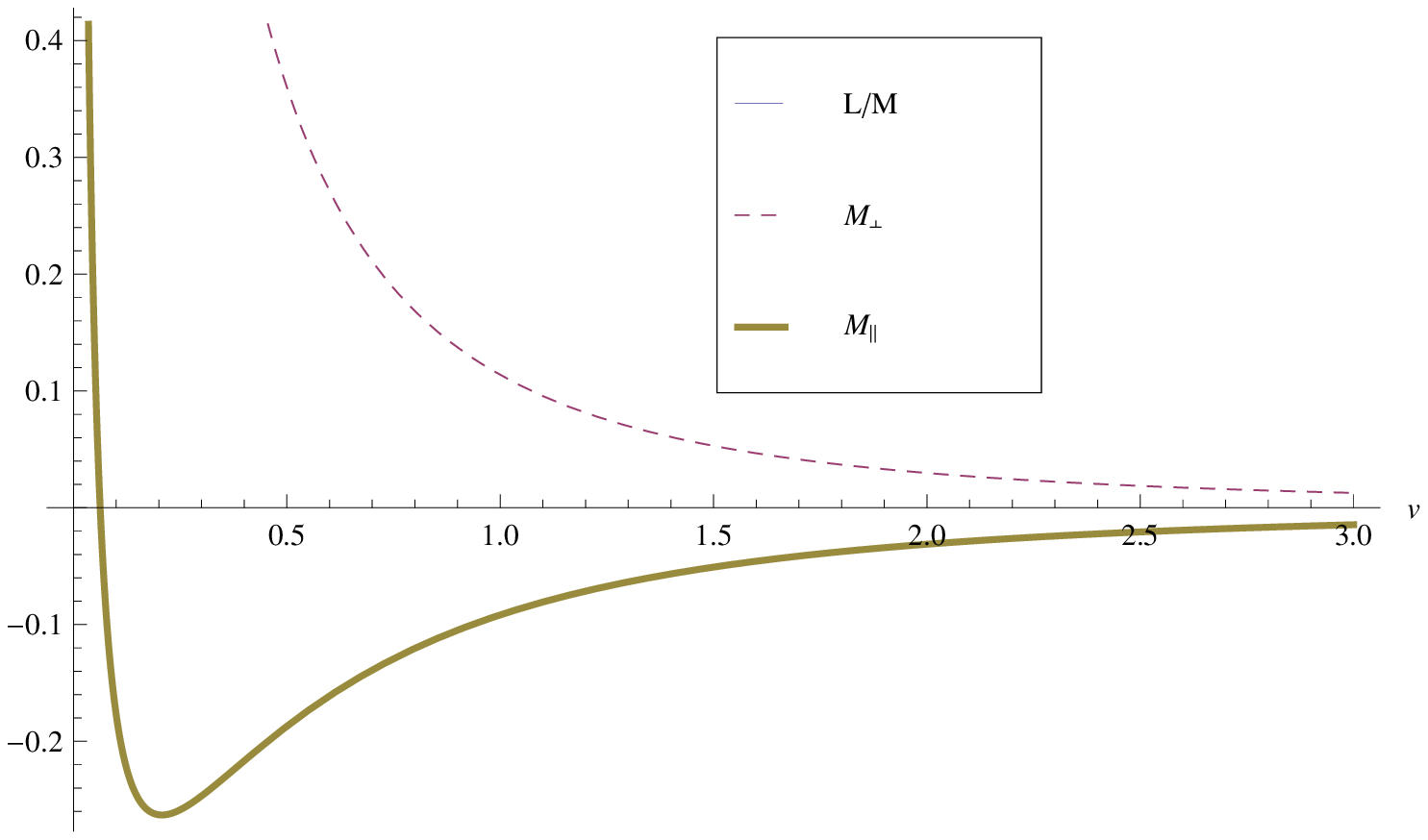}\hspace{1cm}
\caption{Left:A simultaneous plot of ${L\over M}$, $M_{||}$ and $M_{\perp}$ at $z=2.1$. Right: the same plot for $z=8$.}\label{fig4}}

\FIGURE[t]{
\includegraphics[width=12cm]{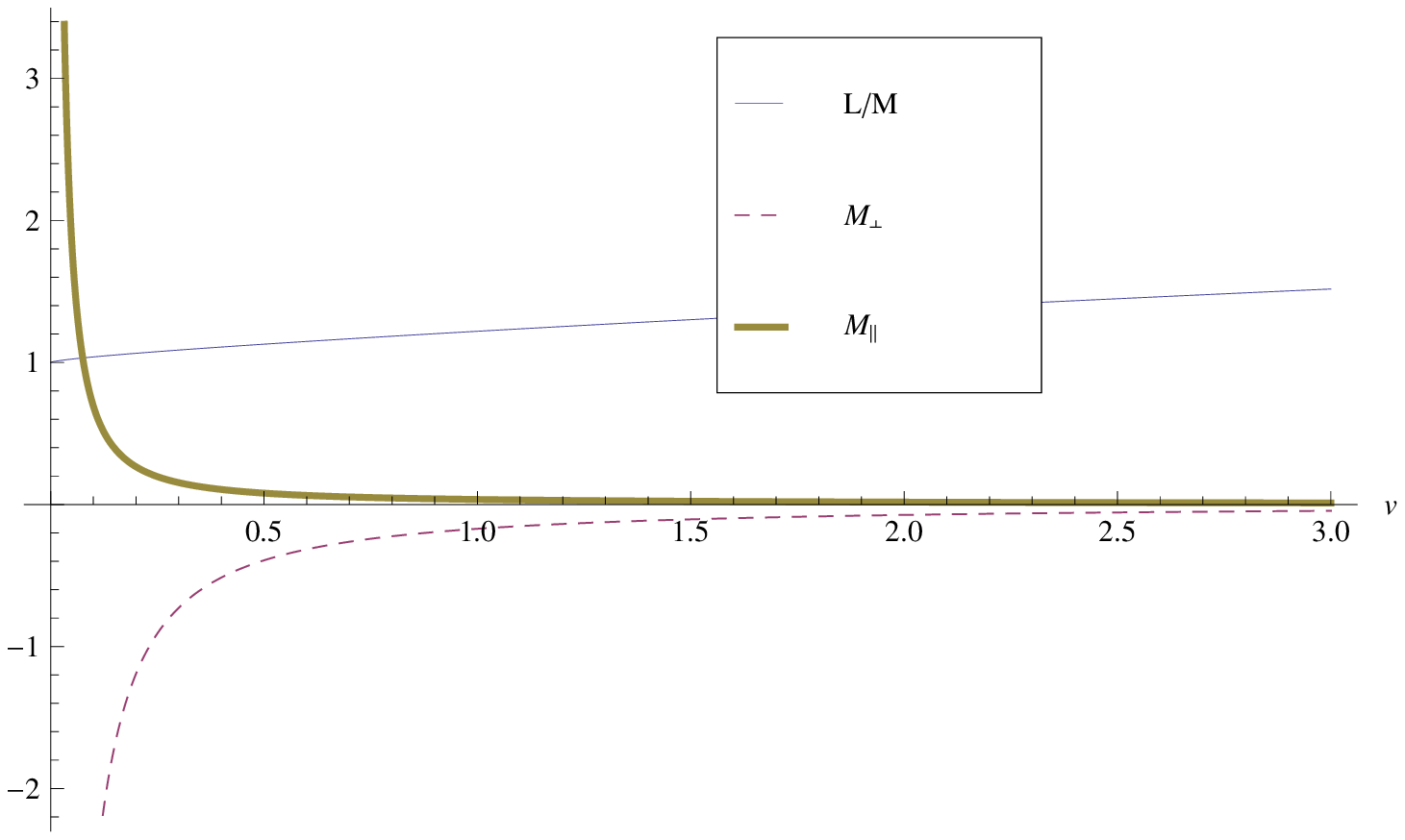}
\caption{Left:A simultaneous plot of ${L\over M}$, $M_{||}$ and $M_{\perp}$ at $\theta=3$, $z={5\over 2}$.}\label{fig5}}

\FIGURE[b]{
\includegraphics[width=7cm]{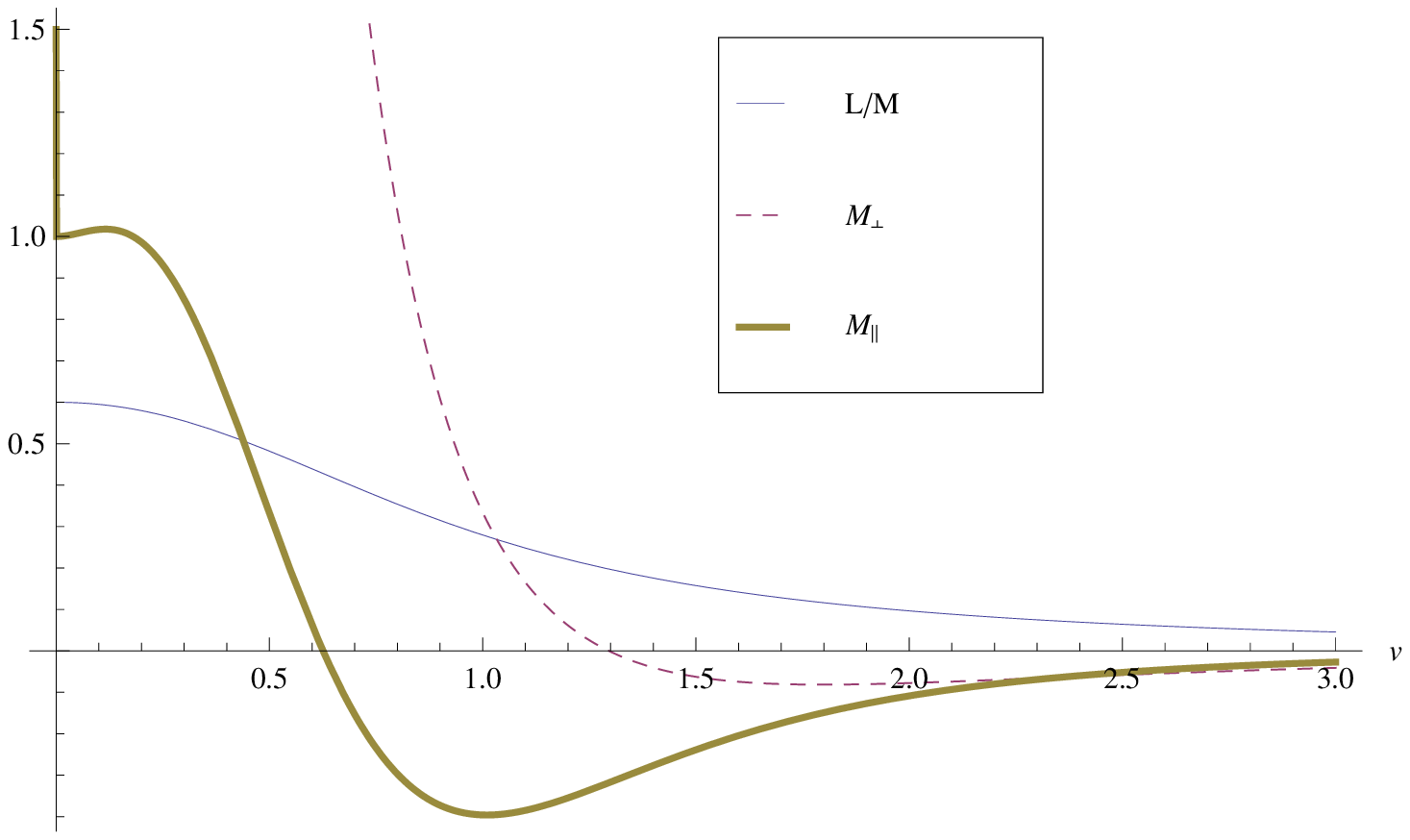}\includegraphics[width=7cm]{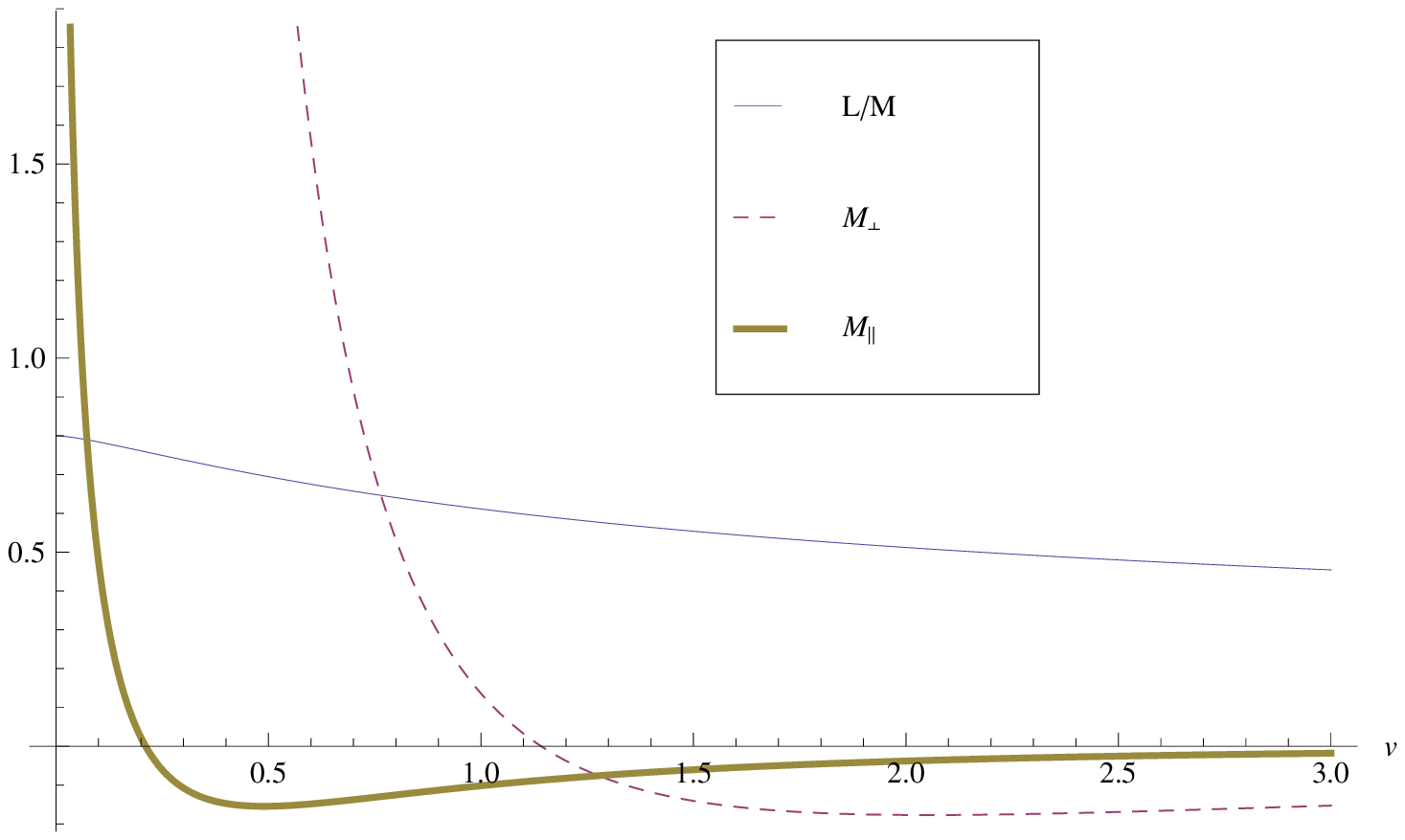}
\caption{Left:A simultaneous plot of ${L\over M}$, $M_{||}$ and $M_{\perp}$ at $\theta=3$, $z={1\over 2}$. Right: A simultaneous plot of ${L\over M}$, $M_{||}$ and $M_{\perp}$ at $\theta=3$, $z=-2$.}\label{fig5a}
}

In figure \ref{fig2} we plot the function ${L_{z,\theta}\over M}=v^{z-\theta\over z-2}F[z,\theta;v^{-{\theta-2\over z-2}}]$ as  function of v, with no hyperscaling violation $\theta=0$, for different values of $z>1$. In all cases $L_{z,\theta}$ asymptotes to a constant as $v\to 0$ while it decreases monotonically and vanishes as $v\to \infty$. Details of the asymptortic expansions can be found at appendix \ref{a}.

The transverse and longitudinal effective masses in (\ref{7.36}) are plotted in figure \ref{fig3} for $1<z<2$, $\theta=0$ and figure \ref{fig4} for $z>2$, $\theta=0$.
As clear from the figures the transverse $M_{\perp}$ becomes negative for sufficiently large velocities in all cases. The turn-around behavior moves to smaller and smaller velocities as $z\to \infty$.
This suggests a potential instability of the system, and in particular that the quadratic fluctuations of the string at large velocities may have negative eigenmodes.
This happens however in a regime where the calculation is unreliable, as we showed in the previous section and a different method is needed to assess the physics.

In figure \ref{fig5} we plot the effective masses in the regime ${\theta-2\over z-2}>0$ and ${z-1\over z-2}>0$. In this regime, $M_{\perp}$ is always negative, while $M_{||}$ always positive.
Here we have a potential instability at all velocities.
Finally in figure \ref{fig5a} we plot the effective masses in the regime ${\theta-2\over z-2}<0$ and ${z-1\over z-2}>0$, with non-trvial positive $\theta$. Here both  $M_{\perp}$
 and $M_{||}$ turn negative at sufficiently high velocities. As we have argued however in the previous section, in this case the trailing string calculation as we have done it
 is unreliable both at high and low velocities.

In  geometries that  are asymptotically AdS in the UV  and flow to one of the generalized Lifshitz geometries in the IR we expect to  have a physically behaved fluctuation problem.

We now proceed to holographically compute the diffusion coefficients following \cite{gkmmn}.

\subsection{The holographic computation of the diffusion constants}

To compute the diffusion coefficients we must compute the real time corelators of the force.  We will use holography to compute them. As the string fluctuations are coupled by definition to the medium force, the holographic computation of the force correlators inbvolves the study of the solution of the string fluctuations to quadratic order according to the standard holographic dictionary.

We therefore consider quadratic fluctuations around the classical string solutions
\be\label{}
 \vec{X}(t,r) = \left( v t + \xi(r) + \delta X^\parl(t,r)\right){\vec{v}\over v} + \delta \vec{X}^\perp(t,r) , \quad \vec{v} \cdot \delta \vec{X}^\perp =0
 \label{7.1}\ee
 \be S_2=-{1\over 2}\int d\tau dr \left[ {\cal G}^{\a\b}_\parl
\partial_{\a}\dx\partial_{\b}\dx + \sum_{i=1}^2 {\cal G}^{\a\b}_\perp \partial_{\a}
\delta X_i^\perp\partial_{\b}\delta X_i^\perp \right]
\label{7.2}\ee where the kinetic operators are defined by

\be {\cal G}^{\a\b}_\perp \equiv {1 \over 2 \pi
\ell_s^2} H^{\a\b}, \qquad {\cal G}^{\a\b}_\parallel \equiv {1 \over 2
\pi \ell_s^2} {H^{\a\b} \over Z^2},  \label{7.3}\ee
with
 \be
 H^{\a\b} = \left( \begin{array}{cc}
-{b^4\over \sqrt{(f-v^2)(b^4f-C^2)}}~~ &  ~~0\\
0~~ & \sqrt{(f-v^2)(b^4f-C^2)}\end{array}\right)
 \label{7.4}\ee
\be
 Z= b^2\sqrt{f-v^2\over b^4f-C^2}=y^{c}\sqrt{1-y^{2a}\over 1-y^{2a+2c}} \sp y={r\over r_s}
 \label{7.5}\ee
and we have defined
\be
a={z-1\over z-2}\sp c={\theta-2\over 2-z}
 \label{7.6}\ee

For a harmonic ansatz of the form
$\delta X^i(r,\tau)=e^{i\omega \tau}\delta X^i(r,\omega)$,
the equations following from the action (\ref{7.2}) are:

\be
\partial_r\left[ R \,\,\partial_r\left(\delta X^{\perp}\right)\right]+{\omega^2b^4\over R}\,\delta X^{\perp}=0,
\label{7.7}\ee
\be
\partial_r\left[{1\over Z^2} R\,\,\partial_r\left(\delta X^{\parallel}\right)\right]+{\omega^2b^4\over Z^2 R}\delta X^{\parallel}=0,
\label{7.8}\ee
where
\be\label{7.9}
R \equiv \sqrt{(f-v^2)(b^4f-C^2)}=f_0\left({r_s\over \ell}\right)^{2a+c}\sqrt{\left(1-y^{2a}\right)\left(1-y^{2a+2c}\right)}
\ee
Changing variables to $y$ in (\ref{7.5}) the equations become
\be
\partial_y\left[\hat R \partial_y\left(\delta X^{\perp}\right)\right]
+{{\cal W}^2 y^{2c}\over \hat R}\delta X^{\perp}=0
 \label{7.10}\ee
 \be
\partial_y\left[{1\over Z^2} R\,\,\partial_y\left(\delta X^{\parallel}\right)\right]+{{\cal W}^2 y^{2c}\over Z^2 R}\delta X^{\parallel}=0,
\label{7.11}\ee

with
\be
{\cal W}={\ell \omega\over f_0}\left({\ell\over r_s}\right)^{z\over z-2}\sp \hat R= \sqrt{\left(1-y^{2a}\right)\left(1-y^{2a+2c}\right)}
 \label{7.12}\ee

As shown in \cite{gkmmn,dress} the transport coefficients, linear in $\omega$ are given by
\be
ImG^{\perp}_R\simeq \chi^{\perp}\omega+\cdots\sp \chi^{\perp}={b^2(r_s)\over 2\pi\ell_s^2}={1\over 2\pi\ls^2}\left({v\over \sqrt{f_0}}\right)^{\theta-2\over 1-z}
\label{7.23}\ee
\be
ImG^{||}_R\simeq \chi^{||}\omega+\cdots\sp \chi^{||}={b^2(r_s)\over 2\pi\ell_s^2Z(r_s)^2}={a+c\over a}\chi^{\perp}
\label{7.24}\ee

Consistency implies that $\chi^{\perp,||}=\hat\eta^{\perp,||}$ and indeed this is the case by inspection.

The difusion constants can be obtained from the symmetric correlators
as
\be
\kappa^{\perp}=\lim_{\omega\to 0}G_{sym}=-2T_s\lim_{\omega\to 0}{ImG_{R}\over \omega}={f_0\sqrt{(z-1)(z+1-\theta)}\over \pi^2\ell\ls^2|z-2|}\left({v\over \sqrt{f_0}}\right)^{z+2-\theta\over z-1}
\label{7.25}\ee
\be
\kappa^{||}={a+b\over a}\kappa^{\perp}={z+1-\theta\over z-1}\kappa^{\perp}
\label{7.26}\ee

The generalized Einstein relations \cite{gkmmn} are
\be
\kappa^{\perp,||}=2T_s~\chi^{\perp,||}
\label{7.27}\ee
with the emergent temperature given in (\ref{6.23}). Note also that (\ref{7.26}) is compatible with (\ref{fri}).

An important condition is the validity of the Markovian approximation for the Langevin evolution.
This demands that ${1\over \eta}\gg {1\over T_s}$. Using the results of the previous sections we obtain that this is valid if
$v^{2-z-\theta\over z-1}$ is sufficiently small. For example at $\theta=0$, the Markovian approximation
 breaks down at high velocities for $1<z<2$, and at low velocities for $z>2$. In such cases the full correlator should be used and not its long time limit.
 In figure \ref{fig10} the allowed regions in the $(z,\theta)$ plane are plotted together with the regions where the  exponent ${2-z-\theta\over z-1}$ is positive. In these regions, the Markovian Langevin evolution is reliable at sufficiently low velocities.

Our results pertain to scaling LV geometries. In realistic situations we expect that such geometries are IR, intermediate of UV asymptotics of a medium.
The general evolution is a combination of such regimes.
The ensemble associated with this type of diffusion is here thermal due to the presence of the world-sheet horizon. This is not always the case and other types of
media or motions may give rise to non thermal ensembles as in \cite{pavlo}.

The fluctuations of a particle trajectory in a LV strongly coupled medium may be associated with Hawking radiation. This has been argued in \cite{haw} for the example of an AdS-Schwartschild black hole. However here there is no regular bulk horizon. In the Lifshitz examples the curvature is regular but there are milder singularities associated to diverging geodesics.
There is no known analogue of Hawking radiation here, but our computation indicates that there may be.

\section{Outlook}

In this paper we examined several questions pertaining to Lorentz Violation in quantum field theory and its interplay with gravity.

Our main points are:
\begin{itemize}

\item LV is intimately interlaced with gravity. LV couplings in QFT are fields in a gravitational sector. Diffeomorphism invariance is intact, and the LV couplings transform under
coordinate/frame changes.

\item Searching for LV is one of the most sensitive ways of looking for new physics: new interactions or modifications of known ones.

\item Energy dissipation/Cerenkov radiation is a generic feature of LV.

\item A general computation can be done in strongly coupled theories with gravity duals. The energy dissipation rate depends non-trivially on two characteristic exponents, the Lifshitz exponent and the hyperscaling-violation exponent.

\end{itemize}

In section \ref{que} we posed several questions, that we will review now.

\begin{enumerate}

\item  {\it How is LV  compatible with diffeomorphism invariance and gravity as we know it?  Is gravity coupled to a LV QFT a consistent theory?}
We have argued on general grounds, that diffeomorphism invariance is not broken. In turn, LV is due to non-trivial backround fields in a gravitational sector.

\item {\it If LI is violated how do we change coordinate systems?}.
The answer to the previous question provides an asnwer here. The coordinate transformations are as usual, LV couplings transform as tensors and Einstein reigns supreme.

\item  {\it Can LI be spontaneously (dynamically) broken?} We have not attempted to answer this question in this paper. However, it remains a very interesting question and there are concrete strategies to attack the problem at weak and strong coupling. We plan to return to this in a future publication.

\item  {\it Is Lorentz invariance an accident of low energies?}. Our analysis indicates that the question is not posed in  the optimal way.
The best way of rephrasing the question is ``How LV couplings renormalize in QFT" and is related to the next question.
The answer to that question is of course not simple, and mostly open. Appart from some examples that have been analysed in the literature
we have added here examples that can be computed using holographic ideas. We have presented some simple ones in section  \ref{braneLV}.

\item  {\it Under what conditions the breaking of LI is ``natural" in the standard sense of Quantum field theory?}
This question is related to the reformulation of the first question. We have given a disussion of the most relevant examples of operators that break LI (vectors and two-index tensors).
However the question remains open in its full breadth.

\item {\it Does the effective speed of light $c_{eff}$ always decrease during RG flow?} This question is correlated with question 4. LI fixed-point QFTs (CFTs) have finite values for the speed of light. Lifshitz QFTs with $z>1$ have an infinite speed of light. Therefore the rate of flow of the speed of light is determined by the nature of the UV and IR fixed point.
We have argued that under very general assumptions (and provided that the answer to question 3 is negative) that LV is due to non-trivial background fields due to charge and energy densities and external EM-like fields. If that is the case, the effects of such non-trivial states/densities at arbitrarily high energy must be negligible.
Therefore the ultimate theory in the UV must be Lorentz Invariant. This does not prohibit however intermediate scaling regions, that have Lifshitz scaling and acts as approximate UV regimes of a LI IR theory. Finally, in theories with a holographic dual, we gave a general formula for the RG flow of the speed of light in terms of two critical exponents,
the Lishitz exponent $z$ and the hyperscaling-violation exponent $\theta$.

\item
{\it Is there a general connection between LV and energy dissipation?}{\it Is there a difference between energy dissipation at strong coupling and weak coupling?}
It is clear that in a gappless theory (or a theory of massless particles like photons, gravitons and others), once LI is broken, and interactions exist, particles with finite energy can always decay to particles with lower energy. We expect therefore that the analogue of Cerenkov radiation is omnipresent. What can vary is the energy dissipation rate that depends on the type of interactions.
We have analyzed a general form of energy dissipation in strongly coupled theories with a gravity dual, using the well known mechanism of energy loss studied in the AdS context.
By considering general scaling generalized Lifshitz geometries, we have given a general formula for the dissipation rate, in terms of two characteristic critical exponents: the
Lishitz exponent $z$ and the hyperscaling-violation exponent $\theta$. We have shown that such a rate depends crucially on the exponents and can be very slow or exponentially fast in different theories. Our techniques break down at large velocities, and new methods are needed to study dissipation there.

\end{enumerate}

We conclude that Lorentz violation is the probably the best window to IR sensitive physics, and both its theoretical and experimental investigation is a very interesting endeavour.

\addcontentsline{toc}{section}{Acknowledgments}
\section*{Acknowledgments}
We would like to thank D. Autiero, L. Blanchet, L. Brink, S. Katsanevas, C. Kounnas, N. Mavromatos, V. Niarchos, F. Nitti,
 A. Petkou, V. Rubakov, A. Schwimmer, S. Sibiryakov, A. Strumia, A. Vikman   for discussions and input. Extra thanks to
  S. Katsanevas and C. Kounnas for a careful reading of the manuscript and comments.

The research presented in this paper has been partially supported by grants  FP7-REGPOT-2008-1-CreteHEPCosmo-228644,
 PERG07-GA-2010-268246, the EU program ``Thales'' ESF/NSRF 2007-2013.
 It has also been co-financed by the European Union (European Social Fund – ESF) and Greek national funds through the
 Operational Program "Education and Lifelong Learning" of the National Strategic Reference Framework (NSRF) under
  "Funding of proposals that have received a positive evaluation in the 3rd and 4th Call of ERC Grant Schemes".

\newpage
 \addcontentsline{toc}{section}{Appendices}
  \renewcommand{\theequation}{\thesection.\arabic{equation}}
\appendix
\section*{Appendix}

\FIGURE[b]{
\includegraphics[width=6cm]{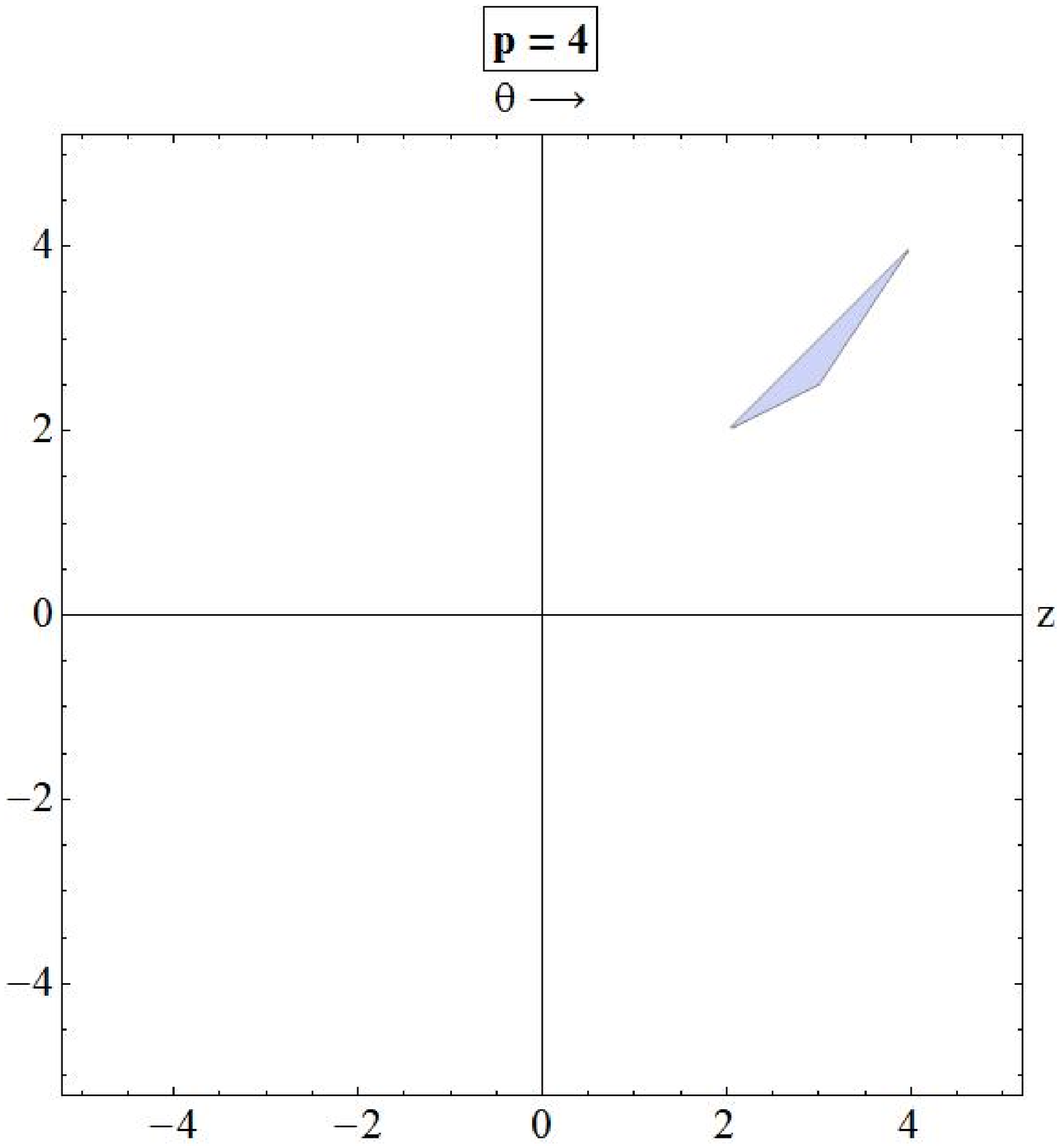}\includegraphics[width=6cm]{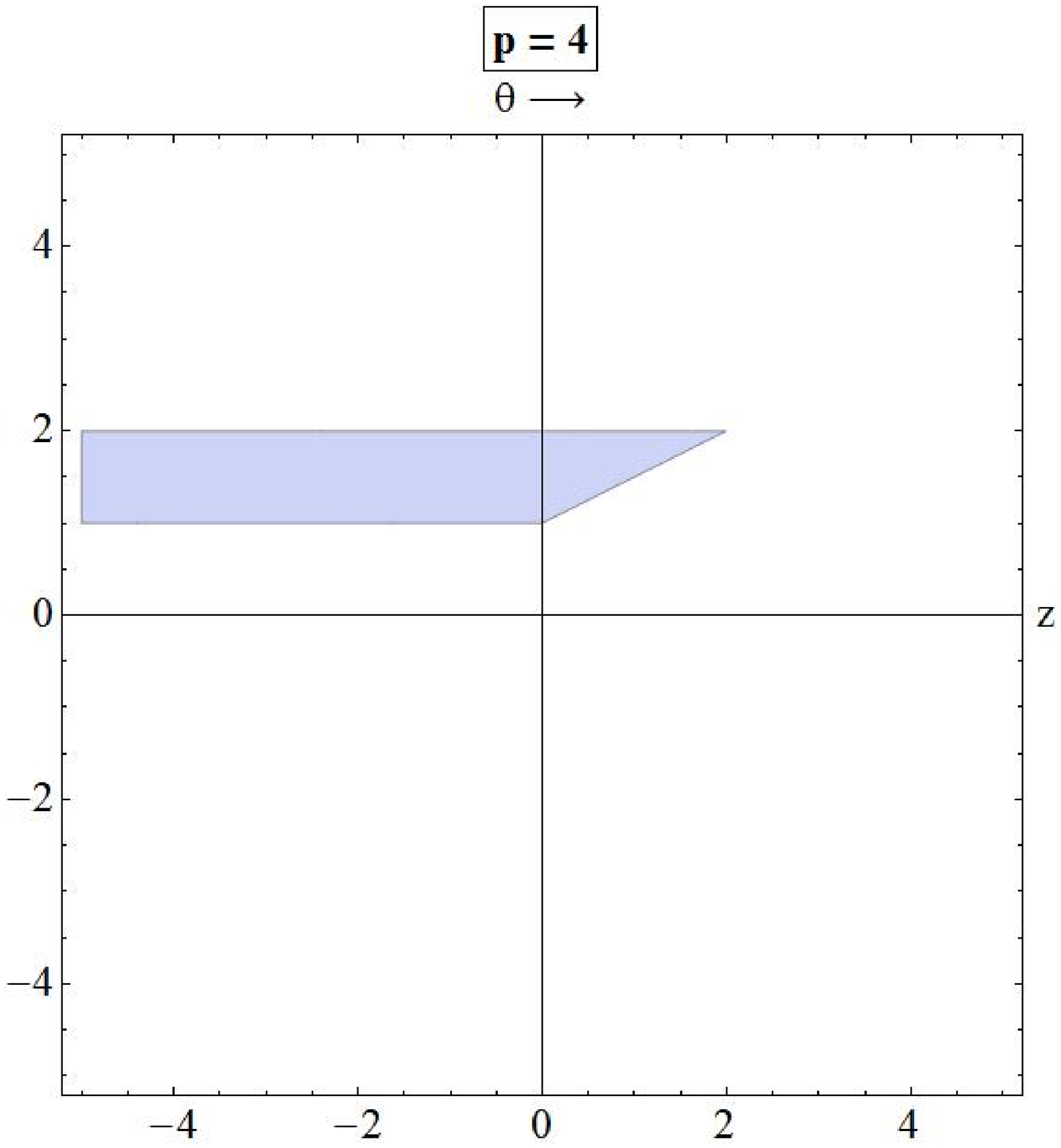}
\caption{Left:generalized Lifshitz geometries in the $(z,\theta)$ plane with ${\theta-2\over z-2}>1$ and ${z-1\over z-2}>0$. Right: generalized Lifshitz geometries in the
$(z,\theta)$ plane with ${\theta-2\over z-2}<{1\over 2}$ kai ${z-1\over z-2}<0$. }\label{fig7}}

\section{The particle action and energy\label{a}}

We will start again from the Nambu-Goto action (\ref{6.6})
\begin{equation}
S_0 = -\frac{1}{2\pi \ls^2} \int dt dr ~ b^2(r)\sqrt{1 - {v^2\over f(r)}+ f(r) \xi^{'2}(r)} \;.
\label{a1.1}\end{equation}
and we will evaluate it in the classical solution (\ref{6.8}) to obtain
\be
S_0=-
\frac{1}{2\pi \ls^2} \int dt\int_{r_b}^{r_h} dr ~ b^4(r)\sqrt{f-v^2\over b^4f-C^2} \;.
\label{a1.2a}\ee
where $r_b$ is a shifted boundary coordinate, while $r_h$ is the position of the bulk horizon.
From this we can read the Lagrangian for the strongly interacting particle as
\be
L=\frac{1}{2\pi \ls^2} \int_{r_b}^{r_h} dr ~ b^4(r)\sqrt{f-v^2\over b^4f-C^2}
\label{a1.3a}\ee
In AdS$_5$ case with $b(r)={\ell\over r}$, $f(r)=1-(\pi Tr)^4$, we obtain
\be
L_{AdS}=\frac{\ell^2\sqrt{1-v^2}}{2\pi \ls^2} \int_{r_b}^{\infty} {dr\over r^2}=M\sqrt{1-v^2}\sp M\equiv {\ell^2\over 2\pi\ell_s^2 r_b}={\sqrt{\lambda}\over 2\pi r_b}
\label{a1.4a}\ee
We now consider the generalized scaling solutions (\ref{6.20}).
The Gubser conditions become,
\be
{2z+3(2-\theta)\over 2(z-1)-\theta}>0\sp {z-1\over 2(z-1)-\theta}>0\sp {2(z-1)+3(2-\theta)\over 2(z-1)-\theta}>0
\label{a1.5}\ee
while thermodynamic stability implies that
\be
{z\over  2(z-1)-\theta}>0
\label{a1.6}\ee
and (\ref{a1.5}) together with (\ref{a1.6}) imply that $z(z-1)>0$.

For the generalized Lifshitz geometries we obtain several cases

\begin{enumerate}

\item  {\bf ${\theta-2\over z-2}> 0$}. In this case the boundary is at $r=0$, and we obtain
\be
L_{\theta,z}(v)=\frac{1}{2\pi \ls^2}\left({r_s\over \ell}\right)^{\theta-2\over 2-z}\int_{r_b}^{\infty
} dr \left({r\over r_s}\right)^{-{2(\theta-2)\over z-2}}\sqrt{
\left({r\over r_s}\right)^{2(z-1)\over z-2}-1\over \left({r\over r_s}\right)^{2(z+1-\theta)\over z-2}-1}\sp {r_s\over \ell}=\left({v^2\over f_0}\right)^{{2-z\over 2(1-z)}}
\label{a1.7}\ee
Since we always have ${z+1-\theta\over z-1}>0$,  the two exponents in the square root have the same sign.
The integral converges at $r=\infty$ when ${\theta-2\over z-2}> 1$.
For large velocities the position of the world-sheet horizon $r_s$ is near the boundary.
In the opposite case of small velocities the horizon is moving deep into the IR.

Changing variables we obtain
\be
L_{\theta,z}(v)=\frac{\ell}{2\pi \ls^2}\left({r_s\over \ell}\right)^{1-{\theta-2\over z-2}}\int_{r_b\over r_s}^{\infty} {d\mu\over  \mu^{{2(\theta-2)\over z-2}}}\sqrt{1-
\mu^{2(z-1)\over z-2}\over 1-\mu^{2(z+1-\theta)\over z-2}}
\label{a1.8}\ee
It seems natural to rescale
\be
v\to \sqrt{f_0}\left({r_b\over \ell}\right)^{z-1\over z-2}~v
\label{a1.8a}\ee
 so that
\be
{r_s\over r_b}=v^{z-2\over z-1}
\label{a1.9}\ee
and rewrite
\be
L_{\theta,z}(v)=M~v^{z-\theta\over z-1}F_+\left[z,\theta;v^{-{z-2\over z-1}}\right]\sp M\equiv \frac{\ell}{2\pi \ls^2}\left({r_b\over \ell}\right)^{1-{\theta-2\over z-2}}
\label{a1.10}\ee
and
\be
F_+\left[z,\theta;x\right]\equiv \int_x^{\infty}{d\mu\over  \mu^{{2(\theta-2)\over z-2}}}\sqrt{1-
\mu^{2(z-1)\over z-2}\over 1-\mu^{2(z+1-\theta)\over z-2}}
\label{a1.11}\ee

We may now evaluate the longitudinal and transverse effective masses,
from
\be
{dL_{\theta,z}(v)\over dv}=\left[{z-\theta\over z-1}L_{z,\theta}+{z-2\over z-1}M\sqrt{1-v^2\over 1-v^{2(z+1-\theta)\over z-1}}\right]{1\over v}
\ee
\be
{d^2L_{\theta,z}(v)\over dv^2}={(z-\theta)(1-\theta)\over (z-1)^2}{L_{z,\theta}\over v^2}+{(z-\theta)(z-2)\over (z-1)^2}{M\over v^2}\sqrt{1-v^2\over 1-v^{2(z+1-\theta)\over z-1}}
+
\ee
$$+{z-2\over z-1}{d\over dv}\left({M\over v}\sqrt{1-v^2\over 1-v^{2(z+1-\theta)\over z-1}}\right)
$$
\be
{dF_+\left[z,\theta;v^{-{z-2\over z-1}}\right]\over dv}={z-2\over z-1}v^{1+\theta-2z\over z-1}\sqrt{1-v^2\over 1-v^{2(z+1-\theta)\over z-1}}
\ee
We obtain
\be
M_{\perp}=-{1\over v}{dL_{\theta,z}(v)\over dv}=-\left[{z-\theta\over z-1}L_{z,\theta}+{z-2\over z-1}M\sqrt{1-v^2\over 1-v^{2(z+1-\theta)\over z-1}}\right]{1\over v^2}\sp M_{||}=
-{d^2L_{\theta,z}(v)\over dv^2}
\ee

To describe the asymptotics of $L_{\theta,z}(v)$ we must distinguish two cases
\begin{enumerate}
\item ${z-1\over z-2}>0$, ${\theta-2\over z-2}> 0$. Here for convergence we must have ${\theta-2\over z-2}>1$. The allowed regions are shown in the left figure \ref{fig7}.
\be
F_+\left[z,\theta;x\right]\simeq {x^{1-{2(\theta-2)\over z-2}}\over 2{\theta-2\over z-2}-1}\left[1+{1\over2}{1-{2(\theta-2)\over z-2}\over 1+{2(z+3-2\theta)\over z-2}}x^{2(z+1-\theta)\over z-2}+\cdots \right]\sp x\to 0
\label{a1.12}\ee
\be
F_+\left[z,\theta;x\right]\simeq {x^{1-{(\theta-2)\over z-2}}\over {\theta-2\over z-2}-1}\left[1-
{1\over 2}{1-{\theta-2\over z-2}\over 1+{2z-\theta\over z-2}}x^{-{2(z-1)\over z-2}}+\cdots\right]
\sp x\to \infty
\label{a1.13}\ee

and we obtain
\be
v\to 0\sp L_{\theta,z}(v)\simeq {M\over {\theta-2\over z-2}-1}\left[1-
{1\over 2}{1-{\theta-2\over z-2}\over 1+{2z-\theta\over z-2}}v^2+\cdots\right]
\label{a1.14}\ee
\be
v\to \infty\sp L_{\theta,z}(v)\simeq{M~ v^{\theta-2\over z-1}\over 2{\theta-2\over z-2}-1}\left[1+{1\over2}{1-{2(\theta-2)\over z-2}\over 1+{2(z+3-2\theta)\over z-2}}v^{-{2(z+1-\theta)\over z-1}}+\cdots\right]
\label{a1.15}\ee

\item ${z-1\over z-2}<0$, ${\theta-2\over z-2}> 0$. The allowed regions are shown in the right figure \ref{fig7}.
In this case
\be
F_+\left[z,\theta;x\right]\equiv \int_x^{\infty}{d\mu\over  \mu^{{(\theta-2)\over z-2}}}\sqrt{1-
\mu^{\Big|{2(z-1)\over z-2}\Big|}\over 1-\mu^{\Big|{2(z+1-\theta)\over z-2}\Big|}}
\label{a1.16}\ee

\be
F_+\left[z,\theta;x\right]\simeq {x^{1-{(\theta-2)\over z-2}}\over {\theta-2\over z-2}-1}\left[1-{1\over2}{1-{(\theta-2)\over z-2}\over 1-{(\theta-2)\over z-2}+\Big|{2(z-1)\over z-2}\Big|}x^{-{2(z-1)\over z-2}}+\cdots\right]\sp x\to 0
\label{a1.17}\ee
\be
F_+\left[z,\theta;x\right]\simeq {x^{1-2{(\theta-2)\over z-2}}\over 2{\theta-2\over z-2}-1}\left[1-
{1\over 2}{2{\theta-2\over z-2}-1\over 1-{2(\theta-z-1)\over z-2}}x^{{2(z-1)\over z-2}}+\cdots\right]
\sp x\to \infty
\label{a1.18}\ee
and we obtain
\be
v\to 0\sp L_{\theta,z}(v)\simeq {M\over {\theta-2\over z-2}-1}\left[1-{1\over2}{1-{(\theta-2)\over z-2}\over 1-{(\theta-2)\over z-2}+\Big|{2(z-1)\over z-2}\Big|}v^{2}+\cdots\right]
\label{a1.19}\ee
\be
v\to \infty\sp L_{\theta,z}(v)\simeq{M~ v^{\theta-2\over z-1}\over 2{\theta-2\over z-2}-1}\left[1-
{1\over 2}{2{\theta-2\over z-2}-1\over 1-{2(\theta-z-1)\over z-2}}{1\over v^2}+\cdots\right]
\label{a1.20}\ee

\end{enumerate}

\FIGURE[t]{
\includegraphics[width=6cm]{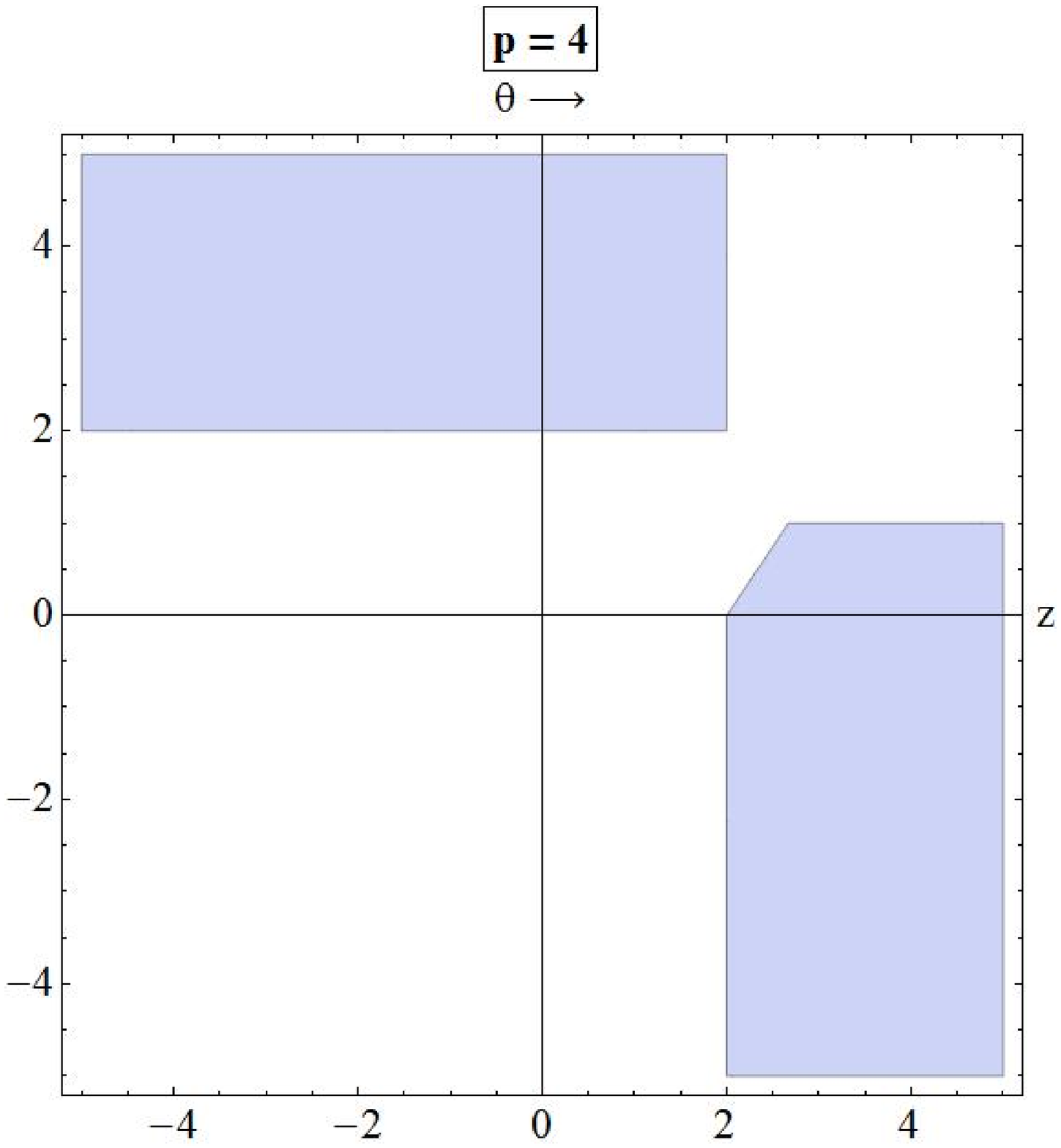}\includegraphics[width=6cm]{plot5a.eps}
\caption{Left:generalized Lifshitz geometries in the $(z,\theta)$ plane with ${\theta-2\over z-2}<0$. All the allowed geometries have ${z-1\over z-2}>0$. }\label{fig8}
}

\FIGURE[b]{
\includegraphics[width=4cm]{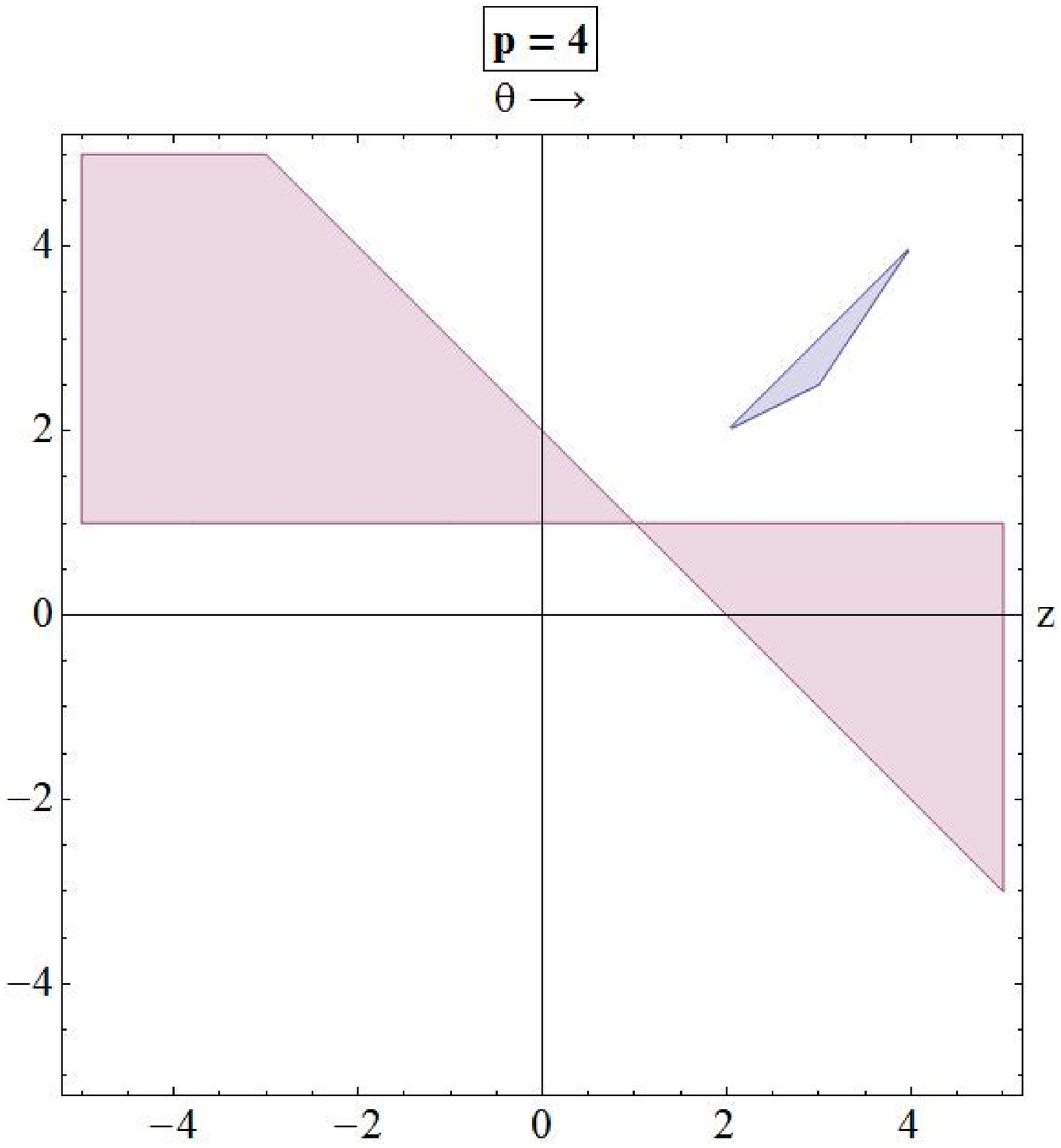}\includegraphics[width=4cm]{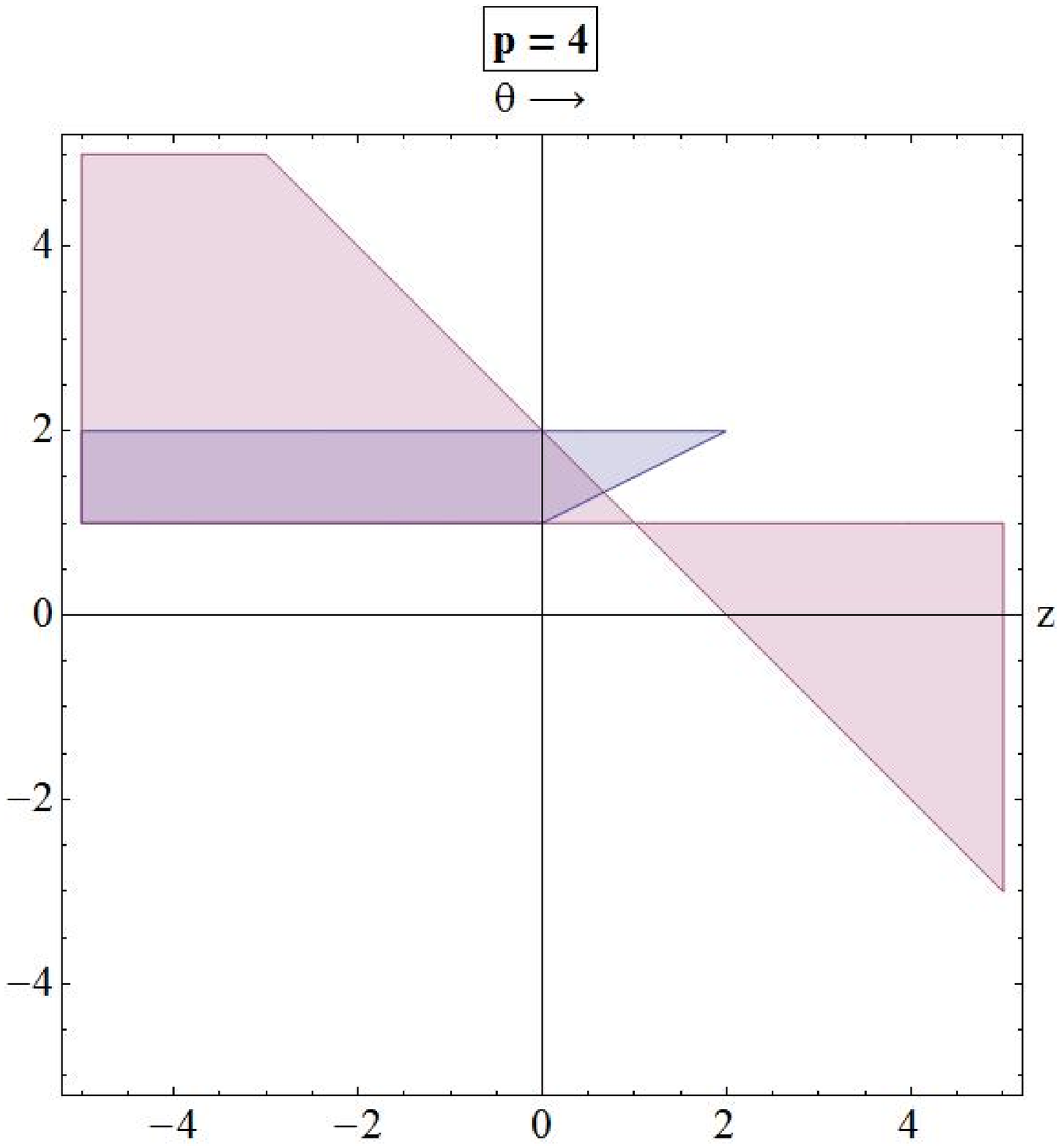}\includegraphics[width=4cm]{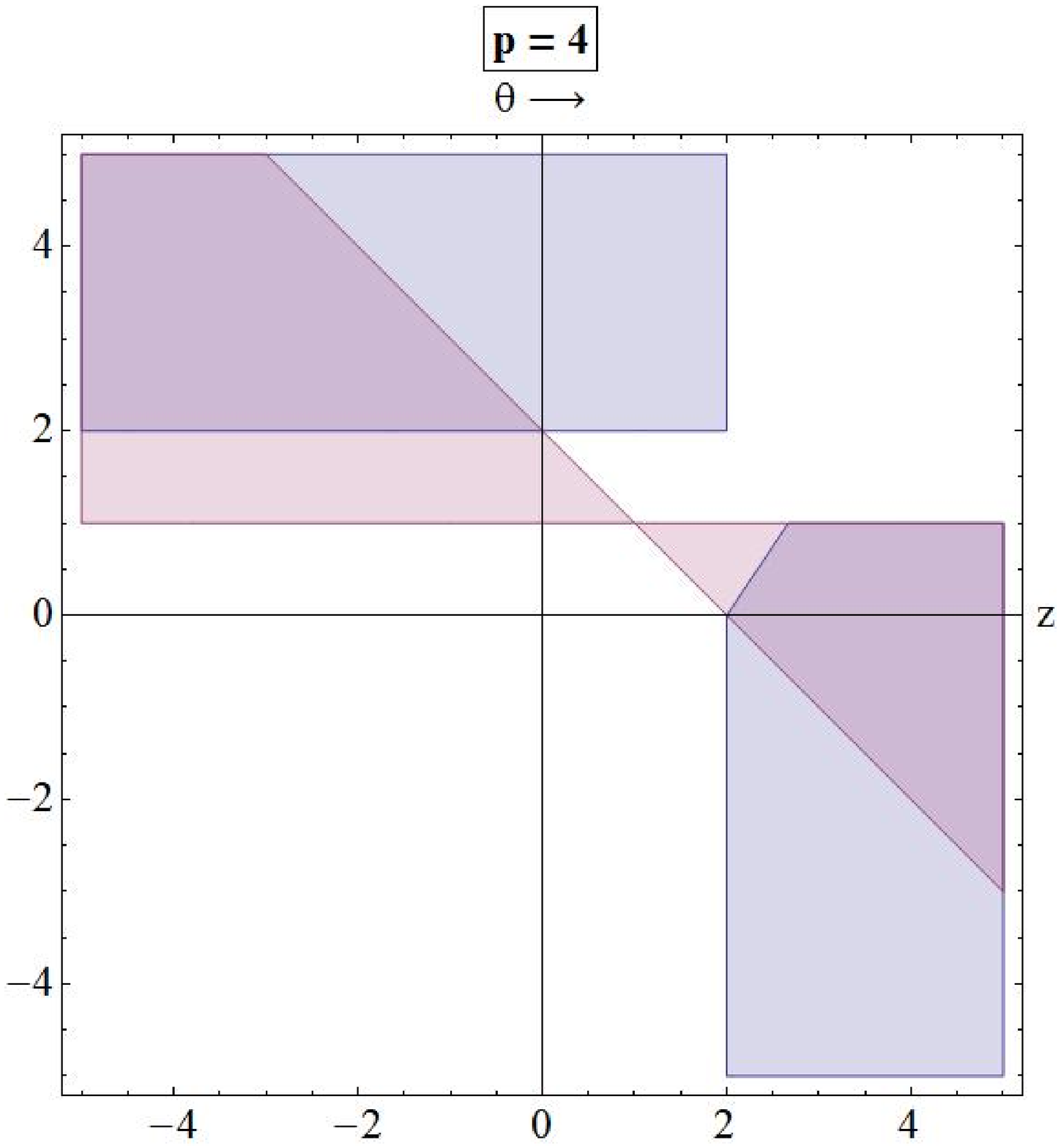}
\caption{Plots of the allowed generalized Lifshitz geometries (blue), and regions where ${(2-z-\theta)\over (z-1)}>0$ (purple).
In the purple allowed regions (overlap blue+purple), the Langevin evolution is Markov at low velocities. In the blue regions only, the process is Markov only at high velocities.
Left figure: The region of case (1a) in appendix \ref{a}. Middle figure: The region of case (2a) in appendix \ref{a}. Right figure: The region of case (2a) in appendix \ref{a}.
 }\label{fig10}}

\item   ${\theta-2\over z-2}< 0$. In this case the boundary is at $r=\infty$, and we obtain
\be
L_{\theta,z}(v)=\frac{1}{2\pi \ls^2}\left({r_s\over \ell}\right)^{\theta-2\over 2-z}\int_{0}^{r_b} dr \left({r\over r_s}\right)^{-{2(\theta-2)\over z-2}}\sqrt{
\left({r\over r_s}\right)^{2(z-1)\over z-2}-1\over \left({r\over r_s}\right)^{2(z+1-\theta)\over z-2}-1}
\label{a1.21}\ee
Again large velocities push the worlsheet horizon near the boundary.

Changing variables we write
\be
L_{\theta,z}(v)=M~v^{z-\theta\over z-1}F_-\left[z,\theta;v^{-{z-2\over z-1}}\right]\sp M\equiv \frac{\ell}{2\pi \ls^2}\left({r_b\over \ell}\right)^{1-{\theta-2\over z-2}}
\label{a1.22}\ee
and
\be
F_-\left[z,\theta;x\right]\equiv \int_0^x{d\mu\over  \mu^{{2(\theta-2)\over z-2}}}\sqrt{1-
\mu^{2(z-1)\over z-2}\over 1-\mu^{2(z+1-\theta)\over z-2}}
\label{a1.23}\ee

Here we obtain
\be
{dF_-\left[z,\theta;v^{-{z-2\over z-1}}\right]\over dv}=-{z-2\over z-1}v^{1+\theta-2z\over z-1}\sqrt{1-v^2\over 1-v^{2(z+1-\theta)\over z-1}}
\ee
\be
{dL_{\theta,z}(v)\over dv}=\left[{z-\theta\over z-1}L_{\theta,z}(v)-{z-2\over z-1}M\sqrt{1-v^2\over 1-v^{2(z+1-\theta)\over z-1}}\right]{1\over v}
\ee
\be
{d^2L_{\theta,z}(v)\over dv^2}={(z-\theta)(1-\theta)\over (z-1)^2}{L_{z,\theta}\over v^2}-{(z-\theta)(z-2)\over (z-1)^2}{M\over v^2}\sqrt{1-v^2\over 1-v^{2(z+1-\theta)\over z-1}}~~-
\ee
$$
-{z-2\over z-1}{d\over dv}{M\over v}\sqrt{1-v^2\over 1-v^{2(z+1-\theta)\over z-1}}
$$
To describe the asymptotics of $L_{\theta,z}(v)$ we must distinguish two cases
\begin{enumerate}
\item ${z-1\over z-2}>0$, ${\theta-2\over z-2}< 0$. The allowed regions are shown in  figure \ref{fig8}.
\be
F_-\left[z,\theta;x\right]\simeq {x^{1-{2(\theta-2)\over z-2}}\over 1-2{\theta-2\over z-2}}\left[1+{1\over2}{1-{2(\theta-2)\over z-2}\over 1+{2(z+3-2\theta)\over z-2}}x^{2(z+1-\theta)\over z-2}+\cdots \right]\sp x\to 0
\label{a1.24}\ee
\be
F_-\left[z,\theta;x\right]\simeq {x^{1-{(\theta-2)\over z-2}}\over 1-{\theta-2\over z-2}}\left[1-
{1\over 2}{1-{\theta-2\over z-2}\over 1+{2z-\theta\over z-2}}x^{-{2(z-1)\over z-2}}+\cdots\right]
\sp x\to \infty
\label{a1.25}\ee
and we obtain
\be
v\to 0\sp L_{\theta,z}(v)\simeq {M\over 1-{\theta-2\over z-2}}\left[1-
{1\over 2}{1-{\theta-2\over z-2}\over 1+{2z-\theta\over z-2}}v^2+\cdots\right]
\label{a1.26}\ee
\be
v\to \infty\sp L_{\theta,z}(v)\simeq{M~ v^{\theta-2\over z-1}\over1- 2{\theta-2\over z-2}}\left[1+{1\over2}{1-{2(\theta-2)\over z-2}\over 1+{2(z+3-2\theta)\over z-2}}v^{-{2(z+1-\theta)\over z-1}}+\cdots\right]
\label{a1.27}\ee

\item ${z-1\over z-2}<0$, ${\theta-2\over z-2}< 0$.
In this case
\be
F_-\left[z,\theta;x\right]\equiv \int_0^{x}{d\mu\over  \mu^{{(\theta-2)\over z-2}}}\sqrt{1-
\mu^{\Big|{2(z-1)\over z-2}\Big|}\over 1-\mu^{\Big|{2(z+1-\theta)\over z-2}\Big|}}
\label{a1.28}\ee

\be
F_-\left[z,\theta;x\right]\simeq {x^{1-{(\theta-2)\over z-2}}\over 1-{\theta-2\over z-2}}\left[1+{1\over2}{1-{(\theta-2)\over z-2}\over 1+{2(z+3-2\theta)\over z-2}}x^{-{2(z+1-\theta)\over z-2}}+\cdots\right]\sp x\to 0
\label{a1.29}\ee
\be
F_-\left[z,\theta;x\right]\simeq {x^{1-2{(\theta-2)\over z-2}}\over 1-2{\theta-2\over z-2}}\left[1-
{1\over 2}{2{\theta-2\over z-2}-1\over 1-{2(\theta-z-1)\over z-2}}x^{{2(z-1)\over z-2}}+\cdots\right]
\sp x\to \infty
\label{a1.30}\ee

and we obtain
\be
v\to 0\sp L_{\theta,z}(v)\simeq {M\over 1-{\theta-2\over z-2}}\left[1+{1\over2}{1-{(\theta-2)\over z-2}\over 1+{2(z+3-2\theta)\over z-2}}v^{2(z+1-\theta)\over z-1}+\cdots\right]
\label{a1.31}\ee
\be
v\to \infty\sp L_{\theta,z}(v)\simeq{M~ v^{\theta-2\over z-1}\over 1-2{\theta-2\over z-2}}\left[1-
{1\over 2}{2{\theta-2\over z-2}-1\over 1-{2(\theta-z-1)\over z-2}}{1\over v^2}+\cdots\right]
\label{a1.32}\ee

There are no acceptable geometries in this class that survive the Gubser bounds in (\ref{a1.5}).

\end{enumerate}
 \end{enumerate}

In all relevant cases above (1a,1b,2a), the action asymptotes to a constant ${M\over \Big|1-{\theta-2\over z-2}\Big|}$ at zero velocity and goes to zero or diverges as $v^{\theta-2\over z-1}$ at large velocities.

The metrics in (\ref{6.20}) can be generalized to include finite temperature  as
\be
ds^2\sim \left({r\over \ell}\right)^{\theta-2\over 2-z}\left[-f(r)dt^2+{dr^2\over f(r)}+dx^idx^i\right]\sp f(r)= f_0 \left({r\over \ell}\right)^{2{1-z\over 2-z}}h\sp
h=1-\left({r\over r_0}\right)^{z+{3\over 2}(2-\theta)\over 2-z}
 \label{6.20a}\ee

 In this case, the physics for small velocities remains qualitatively the same. However the presence of a regular horizon in the metric does not introduce an upper bound on the velocity,
 unlike the LI (AdS) case.

\section{The Langevin equation\label{la}}

We consider a heavy particle which, in a first approximation, experiences a
uniform motion across a medium, with constant velocity $v$. Due
to the interactions with the strongly-coupled medium, the actual
trajectory of the particle  is expected to resemble Brownian motion. To
lowest order, the action for the external particle coupled to the
medium  can be assumed, classically, to be of the form:
\be\label{bound action1} S[X(t)] = S_0 + \int d\tau X_\mu(\tau)
{\cal F}^\mu(\tau) \ee
where $S_0$ is the free quark action, and ${\cal F}(\tau)$ depends
only on the medium degrees of freedom, and plays the role of a
driving force (the ``drag" force).

To obtain an equation for the particle trajectory one needs to trace
over the medium  degrees of freedom. If the interaction energies
are small compared with the quark kinetic energy (therefore for a very
heavy quark, and/or for ultra-relativistic propagation speeds),
tracing over the microscopic degrees of freedom of the medium can
be performed in the semiclassical approximation, and the particle
motion can be described by a {\em classical} generalized Langevin
equation for the position $X^i(t)$, of the form:
\be\label{langeq} {\delta S_0 \over \delta X_i(t)} =
\int_{-\infty}^{+\infty} d\tau ~\theta (\tau) C^{ij}(\tau)
X_j(t-\tau) + \xi^i(t) , \qquad i=1,2,3
\ee
Here, $C^{ij}(t)$ is a {\em memory
kernel}, $\theta(\tau)$ is the Heaviside function and $\xi(t)$ is a Gaussian random variable with
time-correlation: \be\label{noise} \langle\xi^i(t)\xi^j(t')\rangle
= A^{ij} (t-t') \ee

The functions  $A^{ij} (t)$ and $C^{ij}(t)$ are determined
 by the symmetrized and anti-symmetrized real-time correlation functions
 of the forces ${\cal F}(t)$ over the statistical ensemble of the medium:
\be\label{correlators1} C^{ij}(t) = G_{asym}^{ij}(t) \equiv -i
\langle\left[{\cal F}^i(t), {\cal F}^j(0) \right]\rangle, \quad
A^{ij} (t) = G_{sym}^{ij}(t) \equiv -{i\over 2}\langle\left\{{\cal
F}^i(t), {\cal F}^j(0) \right\}\rangle . \ee

The results (\ref{langeq}) and (\ref{correlators1}) are very
general, and do not require any particular assumption about the
statistical ensemble that describes the medium (in particular, they
do not require thermal equilibrium). One way to arrive at equation
(\ref{langeq}) is using the double time formalism and the
Feynman-Vernon influence functional \cite{Feyver}. A clear and
detailed presentation can be found in \cite{kleinert}, chapter 18.

The retarded and advanced Green's function are defined by:
\be
G_R^{ij}(t) = \theta(t)C^{ij}(t), \qquad G_A^{ij}(t) =
-\theta(-t)C^{ij}(t),
\ee
 which lead to the relation
\be\label{comm} C^{ij}(t) = G_R^{ij}(t) - G_A^{ij}(t)
\ee
Notice
that the kernel entering the first term on the right in equation
(\ref{langeq}) is the retarded Green's function, $G_R^{ij}(t) =
\theta(t) C^{ij}(t)$.

It is customary to introduce a {\em spectral density}
$\rho^{ij}(\omega)$ as the Fourier transform of the
anti-symmetrized (retarded) correlator,
\be C^{ij}(t) =
-i\int_{-\infty}^{+\infty} d \omega\, \rho^{ij}(\omega)
e^{-i\omega t}, \qquad G_R^{ij}(\omega) = \int_{-\infty}^{+\infty}
d \omega' \, { \rho^{ij}(\omega') \over \omega - \omega' +
i\epsilon}. \ee

{}From equation (\ref{comm}) and the reality condition $G_A(t) =
G_R(-t)$, or in Fourier space, $G_A(\omega) = G_R^*(\omega)$, we
can relate the spectral density to the imaginary part of the
retarded correlator:
\be \rho^{ij}(\omega) = -{1\over \pi} {\rm Im}\,
G_R^{ij}(\omega) \ee

\paragraph{Local limit.}
Suppose the time-correlation functions  vanish for sufficiently large
separation, i.e. for times much larger than a certain correlation time $\tau_c$.
Then, in the limit $t\gg \tau_c$, equation (\ref{langeq}) becomes
a conventional {\em local} Langevin equation, with local friction and white
noise stochastic term. Indeed, in this regime the noise correlator can be
approximated by
\be A^{ij} (t-t') \approx \kappa^{ij}
\delta(t-t'), \qquad t-t' \gg \tau_c.
\ee
This equation defines the Langevin diffusion constants
$\kappa^{ij}$. Similarly, for the friction term, we define the function
$\gamma^{ij}(t)$ by the relation:
\be C^{ij}(t) = {d\over d t}
\gamma^{ij}(t) \ee
 so that the friction term can be approximated, for large times,
 as: \be\int_0^\infty d\tau C^{ij}(\tau)
X_j(t-\tau) \approx \left(\int_0^\infty d\tau\,
\gamma^{ij}(\tau)\right)\dot{X_j}(t), \qquad t\gg \tau_c. \ee

In this regime, equation (\ref{langeq})
becomes the local Langevin equation with white noise,
\be\label{langeq22} {\delta S_0 \over \delta X_i(t)} + \eta^{ij}
\dot{X}_j(t) = \xi^i(t), \qquad \langle\xi^i(t)\xi^j(t')\rangle =
\kappa^{ij} \delta(t-t'),
\ee
with the self-diffusion and
friction coefficients given by:
\be \label{coeff1} \kappa^{ij} =
\lim_{\omega \to 0} G_{sym}^{ij}(\omega); \qquad \; \eta^{ij}
\equiv \int_0^\infty d\tau\, \gamma^{ij}(\tau) = -\lim_{\omega
\to 0} { {\rm Im}\, G_R^{ij} (\omega) \over \omega}. \ee

In the case of a system at equilibrium with a canonical ensemble
at temperature $T$, one has the following relation between the
Green's functions:
\be\label{thermal1} G_{sym}(\omega) = -
\coth{\omega\over 2T}\, {\rm Im} \,G_R(\omega), \ee
 which using equation
(\ref{coeff1}) leads to the Einstein relation $\kappa^{ij}= 2 T
\eta^{ij}$.
For such a thermal ensemble, the real-time correlators decay exponentially with a scale
set by the inverse temperature, therefore
the typical correlation time is $\tau_c \sim 1/T$.

\end{document}